\documentclass[12pt]{article}

\usepackage{axodraw}

\makeatletter
\if@twoside \oddsidemargin -17pt \evensidemargin 00pt
\else \oddsidemargin 00pt \evensidemargin 00pt
\fi
\expandafter\ifx\csname mathrm\endcsname\relax\def\mathrm#1{{\rm #1}}\fi

\renewcommand{\theequation}{\thesection.\arabic{equation}}
\newcounter{saveeqn}

\@addtoreset{equation}{section}

\makeatother

\def\beq{\begin{equation}}
\def\eeq{\end{equation}}
\def\beqar{\begin{eqnarray}}
\def\eeqar{\end{eqnarray}}
\def\barr#1{\begin{array}{#1}}
\def\earr{\end{array}}
\def\bfi{\begin{figure}}
\def\efi{\end{figure}}
\def\btab{\begin{table}}
\def\etab{\end{table}}
\def\bce{\begin{center}}
\def\ece{\end{center}}
\def\nn{\nonumber}
\def\disp{\displaystyle}
\def\text{\textstyle}

\def\veps{\varepsilon}

\def\refeq#1{\mbox{(\ref{#1})}}
\def\reffi#1{\mbox{Fig.~\ref{#1}}}
\def\reffis#1{\mbox{Figs.~\ref{#1}}}
\def\reftas#1{\mbox{Tables~\ref{#1}}}
\def\refse#1{\mbox{Sect.~\ref{#1}}}

\def\citere#1{\mbox{Ref.~\cite{#1}}}
\def\citeres#1{\mbox{Refs.~\cite{#1}}}

\def\mathswitchr#1{\relax\ifmmode{\mathrm{#1}}\else$\mathrm{#1}$\fi}

\newcommand{\PS}{\mathswitchr S}
\newcommand{\PV}{\mathswitchr V}
\newcommand{\PW}{\mathswitchr W}
\newcommand{\PZ}{\mathswitchr Z}

\newcommand{\PH}{\mathswitchr H}
\newcommand{\Pe}{\mathswitchr e}

\newcommand{\Pf}{f}

\newcommand{\Pep}{\mathswitchr {e^+}}
\newcommand{\Pem}{\mathswitchr {e^-}}

\newcommand{\PWm}{\mathswitchr {W^-}}

\def\mathswitch#1{\relax\ifmmode#1\else$#1$\fi}

\newcommand{\Mf}{\mathswitch {m_\Pf}}

\newcommand{\MS}{\mathswitch {M_\PS}}
\newcommand{\MV}{\mathswitch {M_\PV}}
\newcommand{\MW}{\mathswitch {M_\PW}}

\newcommand{\MZ}{\mathswitch {M_\PZ}}

\newcommand{\scrs}{\scriptscriptstyle}
\newcommand{\sw}{\mathswitch {s_{\scrs\PW}}}
\newcommand{\cw}{\mathswitch {c_{\scrs\PW}}}

\def\slash#1{{\setbox0=\hbox{$#1$}
  \rlap{\ifdim\wd0>.7em\kern.22\wd0\else\kern.1\wd0\fi /}#1}}
\newcommand{\as}{a\hspace{-0.5em}/\hspace{0.1em}}
\newcommand{\es}{\varepsilon \hspace{-0.5em}/\hspace{0.1em}}
\newcommand{\ks}{k\hspace{-0.52em}/\hspace{0.1em}}

\newcommand{\ri}{{\mathrm{i}}}

\newcommand{\M}{{\cal {M}}}

\def\sgn{\mathop{\mathrm{sgn}}\nolimits}

\def\draftdate{\relax}
\def\mda{\relax}
\def\mua{\relax}
\def\mla{\relax}
\def\draft{
\def\thtystars{******************************}
\def\sixtystars{\thtystars\thtystars}
\typeout{}
\typeout{\sixtystars**}
\typeout{* Draft mode!
         For final version remove \protect\draft\space in source file *}
\typeout{\sixtystars**}
\typeout{}
\def\draftdate{\today}
\def\mua{\marginpar[\boldmath\hfil$\uparrow$]%
                   {\boldmath$\uparrow$\hfil}%
                    \typeout{marginpar: $\uparrow$}\ignorespaces}
\def\mda{\marginpar[\boldmath\hfil$\downarrow$]%
                   {\boldmath$\downarrow$\hfil}%
                    \typeout{marginpar: $\downarrow$}\ignorespaces}
\def\mla{\marginpar[\boldmath\hfil$\rightarrow$]%
                   {\boldmath$\leftarrow $\hfil}%
                    \typeout{marginpar: $\leftrightarrow$}\ignorespaces}
\def\Mua{\marginpar[\boldmath\hfil$\Uparrow$]%
                   {\boldmath$\Uparrow$\hfil}%
                    \typeout{marginpar: $\Uparrow$}\ignorespaces}
\def\Mda{\marginpar[\boldmath\hfil$\Downarrow$]%
                   {\boldmath$\Downarrow$\hfil}%
                    \typeout{marginpar: $\Downarrow$}\ignorespaces}
\def\Mla{\marginpar[\boldmath\hfil$\Rightarrow$]%
                   {\boldmath$\Leftarrow $\hfil}%
                    \typeout{marginpar: $\Leftrightarrow$}\ignorespaces}
\overfullrule 5pt
\oddsidemargin -15mm
\marginparwidth 29mm
}

\makeatletter

\def\eqnarray{\stepcounter{equation}\let\@currentlabel=\theequation
\global\@eqnswtrue
\global\@eqcnt\z@\tabskip\@centering\let\\=\@eqncr
$$\halign to \displaywidth\bgroup\hskip\@centering
  $\displaystyle\tabskip\z@{##}$\@eqnsel&\global\@eqcnt\@ne
  \hskip 2\arraycolsep \hfil${##}$\hfil
  &\global\@eqcnt\tw@ \hskip 2\arraycolsep $\displaystyle\tabskip\z@{##}$\hfil
   \tabskip\@centering&\llap{##}\tabskip\z@\cr}
\def\appendix{\par
 \setcounter{section}{0} \setcounter{subsection}{0}
 \def\thesection{\Alph{section}}}

\makeatother

\begin{document}

\thispagestyle{empty}
\def\thefootnote{\fnsymbol{footnote}}
\setcounter{footnote}{1}
\null
\draftdate\hfill CERN-TH/98-143 \\
\strut\hfill hep-ph/9805445
\vskip 0cm
\vfill
\begin{center}
{\Large \bf
Weyl--van-der-Waerden formalism \\[.5em]
for helicity amplitudes of massive particles
\par} \vskip 3.5em
{\large
{\sc Stefan Dittmaier%
}\\[1ex]
{\normalsize \it Theory Division, CERN\\
CH-1211 Geneva 23, Switzerland}\\[2ex]
}
\par \vskip 1em
\end{center}\par
\vskip .0cm 
\vfill 
{\bf Abstract:} \par 
The Weyl--van-der-Waerden spinor technique for calculating helicity
amplitudes of massive and massless particles is presented in a form 
that is particularly well suited to a direct implementation in 
computer algebra. 
Moreover, we explain how to exploit discrete symmetries and how to avoid
unphysical poles in amplitudes in practice.
The efficiency of the formalism
is demonstrated by giving explicit compact results for the helicity
amplitudes of the processes $\gamma\gamma\to\Pf\bar\Pf$, 
$\Pf\bar\Pf\to\gamma\gamma\gamma$, $\mu^-\mu^+\to\Pf\bar\Pf\gamma$.
\par
\vskip 1cm
\noindent
CERN-TH/98-143 \\
May 1998 
\par
\null
\setcounter{page}{0}
\clearpage
\def\thefootnote{\arabic{footnote}}
\setcounter{footnote}{0}

\section{Introduction}

The calculation of scattering amplitudes and cross sections in lowest
order of perturbation theory is one of the standard problems of
elementary particle phenomenology in practice. Although this task is 
very simple for $1\to 2$ particle decays and $2\to 2$ particle reactions, 
the situation can become arbitrarily complicated for more particles in the
final state, as the number $N_D$ of contributing Feynman diagrams rapidly
increases. Squaring amplitudes and using completeness relations for the 
wave functions does not represent an appropriate approach if $N_D$ is 
large, because one would get $N_D^2$ contributions. A reasonable way to
avoid an explosion of the algebraic work consists in using spinor
formalisms for the evaluation of amplitudes for definite polarization
configurations. The squaring of the amplitudes and the spin summations
are then performed numerically. Of course, the number of different
polarization configurations also grows for more and more external
particles, but in practice this fact is much less problematic than
squaring amplitudes. In particular, the number of independent matrix
elements can be reduced by exploiting discrete symmetries.
Although the commonly applied spinor formalisms rely on the
four-dimensionality of space-time, they are also useful for the
calculation of higher-order corrections if the singularities in
dimensional regularization are split off and controlled separately.

Various spinor techniques have been proposed and successfully applied 
in recent years. The first versions \cite{sp1} proposed a clever 
choice of photon and gluon polarization vectors, which was dictated 
by the momenta of the attached fermions. This choice forces a lot of
terms in the calculation to vanish, and relatively compact amplitudes 
result. Although the initial
restriction to massless particles could be overcome \cite{sp2}, all
those formalisms have in common that the usual Dirac algebra is still
present. The final results for amplitudes are given in terms of standard 
products $\bar u_1(p_1)u_2(p_2)$ of Dirac spinors. This form is useful
for numerical evaluations, but too involved for further analytical
purposes. 

It is more elegant and transparent to express all needed ingredients, 
i.e.\ momenta, polarization vectors, and Dirac spinors, in terms of a single 
mathematical object. This leads us to two-component Weyl--van-der-Waerden 
(WvdW) spinors \cite{wvdw29}, which form the fundamental representations
of the Lorentz group, and the related spinor calculus (see also
\citere{ca71}). Since WvdW spinors are closely related to light-like vectors,
the application of the WvdW spinor calculus is extremely simple for
massless particles. Such applications can be found, e.g.\ in \citere{fa83}.
The WvdW technique can also be used for massive particles if
time-like momenta are decomposed into two light-like ones. This
procedure was applied in \citeres{be91,bo93} for the calculation of
bremsstrahlung corrections to production processes of massive weak
gauge bosons, leading to rather compact results. 
However, the WvdW spinor formalism for massive particles is not yet well 
documented in the literature. In connection with the actual applications
presented in \citeres{be91,bo93}, the formalism was only briefly
sketched, as far as was relevant for the special cases under investigation.
More complete presentations of the technique for massive particles are
either not suited for direct practical use \cite{an83} or appeared in
unpublished studies \cite{kudis,sddis}.

In addition to the more technical motivation above for considering
amplitudes with definite polarization configurations, one should add
some remarks on the physical role of polarized massive particles.
In contrast with massless particles, the spin orientations of massive 
particles do not transform covariantly under Lorentz transformations;
in particular, the property of being a helicity eigenstate is
frame-dependent. Nevertheless, the polarization states of a massive
particle also carry valuable physical information about the structure
of its interaction. For instance, the longitudinal modes of the massive
weak gauge bosons serve as a window into the scalar sector of the
electroweak theory, and the reconstruction of the top-quark spin from
its decay products reveals information about its static properties and 
form factors. For various aspects of physics with polarized massive
particles, see \citere{ecfa} and references therein.

The aim of this paper is threefold. First, we give an introduction into 
the WvdW spinor technique for massive and massless particles. The
presentation is deliberately held at a basic and detailed level, in
order to facilitate the practical use of the method. In particular, we
present all ingredients needed for an implementation in computer
algebra. As wave functions for massive spin-$\frac{1}{2}$ and spin-1
particles, we take the helicity eigenstates formulated in \citere{sddis}, 
which appear to be very simple. The wave functions and the momenta of the 
corresponding massive particles are composed of the same set of auxiliary 
spinors, leading to simplifications in the calculation of amplitudes. 
Secondly, we show how to exploit the advantages of the
formalism in practice. We explain the use of discrete symmetries, in
order to reduce the algebraic work, and give a prescription for avoiding
unphysical poles in amplitudes. In the presented examples, we treat the
polarizations of massive fermions in a generic way, i.e.\ explicit 
results for definite helicity configurations follow from a few compact  
generic amplitudes upon setting auxiliary spinors to specific values.
Thirdly, we give explicit results for helicity amplitudes of
phenomenologically relevant processes involving massive particles. 
We explicitly treat the reactions $\gamma\gamma\to\Pf\bar\Pf$, 
$\Pf\bar\Pf\to\gamma\gamma\gamma$, $\mu^-\mu^+\to\Pf\bar\Pf\gamma$ in
the framework of the electroweak Standard Model, and show how these 
results can be carried over to other processes such as
$\gamma\gamma\to\Pf\bar\Pf\gamma$, $\Pem\gamma\to\Pem\gamma\gamma$, 
$\Pem\gamma\to\Pem\Pem\Pep$ by using crossing relations. 

The outline of the paper is as follows. In \refse{se:def} we set our
basic conventions for the WvdW spinor calculus and introduce helicity
eigenstates for spin-$\frac{1}{2}$ and spin-1 particles. The necessary
ingredients for calculating helicity amplitudes are described in
\refse{se:helamp}, specifically containing a prescription for
translating Feynman rules into the WvdW formalism, a formulation
of discrete symmetries for helicity amplitudes, and a prescription for
avoiding unphysical poles in amplitudes.
The physical applications are presented in
\refse{se:app}, and \refse{se:sum} contains a summary.

\section{Basic definitions}
\label{se:def}

\subsection{Spinors}

The basic philosophy of the WvdW formalism is to reduce all mathematical
objects that belong to higher-dimensional representations of the
Lorentz group to the two-dimensional irreducible
representations $D(\frac{1}{2},0)$ and $D(0,\frac{1}{2})$.
According to these representations we distinguish covariant and
contravariant WvdW spinors $\psi_{A}$ and $\psi^{\dot A}$,
respectively. Indices of WvdW spinors are denoted by capital letters
throughout. 
The transition between the two-dimensional representations, which are  
non-equivalent, is achieved by complex conjugation and a similarity
transformation. Complex conjugation is consistently indicated by dotting
(undotting) indices, i.e.
\beq
\psi_{\dot A}=(\psi_A)^*, \qquad
\psi^{A}=(\psi^{\dot A})^*. 
\eeq
The similarity transformation is mediated
by the antisymmetric 2$\times$2 matrix $\epsilon=\ri\sigma^2$, where 
$\sigma^a$ ($a=1,2,3$) are the standard Pauli matrices,
\beq
\epsilon^{AB}=\epsilon^{{\dot A}{\dot B}}=
\epsilon_{AB}=\epsilon_{{\dot A}{\dot B}}=
\pmatrix{0&+1\cr -1&0}.
\eeq
The matrix $\epsilon$ defines how to raise and lower spinor
indices,
\beq
\psi^{A}=\epsilon^{AB}\psi_{B}, \qquad
\psi^{\dot A}=\epsilon^{\dot A\dot B}\psi_{\dot B}, \qquad
\psi_{A}=\psi^{B}\epsilon_{BA}, \qquad
\psi_{\dot A}=\psi^{\dot B}\epsilon_{{\dot B}{\dot A}}.
\eeq
Moreover, $\epsilon$ provides the set of Clebsch--Gordan coefficients
that project the product representations 
$D(\frac{1}{2},0)\otimes D(\frac{1}{2},0)$ and
$D(0,\frac{1}{2})\otimes D(0,\frac{1}{2})$ onto the trivial
representation. For this reason, it is possible to define a
Lorentz-invariant spinor product through
\beq
\langle\phi\psi\rangle = \phi_{A}\psi^{A} = 
\phi_1\psi_2-\phi_2\psi_1, \qquad
\langle\phi\psi\rangle^* = \phi_{\dot A}\psi^{\dot A} =
(\phi_1\psi_2-\phi_2\psi_1)^*, 
\eeq
and $\epsilon$ is called {\it spinor metric}. By definition, the spinor
product is antisymmetric:
\beq
\langle\phi\psi\rangle = -\langle\psi\phi\rangle, \qquad
\langle\phi\phi\rangle = 0.
\eeq
The fact that an antisymmetric tensor built of two-dimensional objects
cannot have a rank higher than 2 implies a Schouten identity for the
spinor metric,
\beq
\epsilon^{AB}\epsilon^{CD} + \epsilon^{AC}\epsilon^{DB} +
\epsilon^{AD}\epsilon^{BC} = 0,
\label{eq:schouten}
\eeq
which in terms of spinor products reads
\beq
\langle \phi\psi\rangle\langle \xi\eta\rangle +
\langle \phi\xi\rangle\langle \eta\psi\rangle +
\langle \phi\eta\rangle\langle \psi\xi\rangle = 0.
\label{eq:schouten2}
\eeq

\subsection{4-vectors}
\label{se:4vectors}

Minkowski 4-vectors belong to the representation 
$D(\frac{1}{2},\frac{1}{2}) = D(\frac{1}{2},0) \otimes D(0,\frac{1}{2})$
of the Lorentz group. The transition of the usual form of a
4-vector $k^\mu=(k^0,{\bf k})$ to the
spinor representation $D(\frac{1}{2},\frac{1}{2})$ is provided by
the matrices
\beq
\sigma^{\mu,{\dot A}B}=(\sigma^0,\mbox{\boldmath{$\sigma$}}), \qquad
\sigma^{\mu}_{A{\dot B}} = (\sigma^0,-\mbox{\boldmath{$\sigma$}}), 
\eeq
consisting of the 2-dimensional unit matrix $\sigma^0$
and the Pauli matrices $\sigma^a$. 
Each 4-vector $k^\mu$ is related to a 2$\times$2 matrix
\beq
K_{{\dot A}B} = k^\mu\sigma_{\mu,{\dot A}B} = 
\pmatrix{k^0+k^3&k^1+\ri k^2\cr k^1-\ri k^2&k^0-k^3},
\label{eq:kmatrix}
\eeq
which is Hermitian if the components of $k^\mu$ are real.
The rules for dotting, undotting, raising, and lowering
spinor indices also apply to the indices of the $\sigma$ matrices; 
in particular, we have 
\beq
\sigma^{\mu}_{{\dot A}B} = \sigma^{\mu,{\dot C}D}\,
\epsilon_{{\dot C}{\dot A}}\,\epsilon_{DB},
\qquad
\sigma^{\mu}_{A{\dot B}} = (\sigma^{\mu}_{{\dot A}B})^*.
\eeq
We note that the coefficients of the transpose $K^{\mathrm{T}}$ of a
matrix $K$ read $K_{\dot BA}$ if the ones of $K$ are denoted by
$K_{\dot AB}$, i.e.\ transposing a matrix interchanges the spinor
indices without moving the position of the dot.
Thus, the Hermiticity of the $\sigma$ matrices is expressed by
\beq
\sigma^{\mu,\dot AB} = \sigma^{\mu,B\dot A}, \qquad
\sigma^\mu_{A\dot B} = \sigma^\mu_{\dot BA}.
\eeq
The $\sigma$ matrices obey the important relations
\beq
\sigma^{\mu}_{{\dot A}B}\,\sigma^{\nu,{\dot A}B}=2g^{\mu\nu}, \qquad
\sigma^{\mu}_{{\dot A}B}\,\sigma^{\nu,{\dot A}C} +
\sigma^{\nu}_{{\dot A}B}\,\sigma^{\mu,{\dot A}C}=2g^{\mu\nu}\delta_B^C, \qquad
\sigma^{\mu}_{{\dot A}B}\,\sigma_{\mu,\dot CD}=
2\epsilon_{{\dot A}{\dot C}}\,\epsilon_{BD}.
\eeq
The first of these relations translates 
the Minkowski inner product of two 4-vectors $k^\mu$ and $p^\mu$ into
\beq
2k\cdot p = k_\mu\, 2g^{\mu\nu}\, p_\nu = 
k_\mu\, \sigma^{\mu}_{{\dot A}B}\sigma^{\nu,{\dot A}B}\, p_\nu = 
K_{{\dot A}B} P^{{\dot A}B},
\label{eq:kp}
\eeq
and the second one implies
\beq
K_{{\dot A}B} K^{{\dot A}C} = k^2 \delta_B^C.
\eeq

In order to reduce terms involving a matrix $K_{{\dot A}B}$ to
spinor products, it is necessary to express $K_{{\dot A}B}$ in terms of
spinors. For a real 4-vector the matrix $K_{{\dot A}B}$ is Hermitian and 
can be decomposed into its eigenvectors $n_{i,A}$ $(i=1,2)$ and eigenvalues 
$\lambda_i$:
\beqar
K_{{\dot A}B} &=& 
\sum_{i=1,2}\lambda_i\, n_{i,{\dot A}}\,n_{i,B}, \qquad
\lambda_{1,2} = k^0 \pm \vert{\bf k}\vert,
\nn\\
n_{1,A} &=&
\pmatrix{e^{-\ri\phi}\cos\frac{\theta}{2}\cr\sin\frac{\theta}{2}}, \qquad 
n_{2,A} = 
\pmatrix{\sin\frac{\theta}{2}\cr -e^{+\ri\phi}\cos\frac{\theta}{2}},
\label{eq:momwvdw}
\eeqar
where $\theta$ and $\phi$ denote the polar and azimuthal angles of 
${\bf k}=|{\bf k}|{\bf e}$, respectively,
\beq
{\bf e} = \pmatrix{\cos\phi\sin\theta \cr \sin\phi\sin\theta
\cr \cos\theta}.
\eeq
{}For time-like vectors ($k^2>0$) it is often convenient to include the
eigenvalues $\lambda_i$ in the normalization of the eigenvectors,
resulting in
\beq
K_{{\dot A}B} = \sum_{i=1,2} \kappa_{i,{\dot A}}\,\kappa_{i,B}, \qquad
\kappa_{i,A} = \sqrt{\lambda_i}\, n_{i,A}.
\label{eq:momwvdw2}
\eeq
The phases of the $n_{i,A}$ are chosen such that 
the orthonormality relations read
\beq
\langle n_i n_i\rangle = 0, \qquad
\langle n_2 n_1\rangle = -\langle n_1 n_2\rangle = +1.
\eeq
Note also the relations for the eigenvalues $\lambda_i$:
\beq
{\mathrm{det}}(K_{{\dot A}B}) = \lambda_1\lambda_2 = k^2, \qquad
\langle\kappa_2\kappa_1\rangle = \sqrt{\lambda_1\lambda_2} = \sqrt{k^2}.
\eeq

The special case of a light-like vector $(k^2=0)$ is of
particular importance. In this case the eigenvalue $\lambda_2$ of 
\refeq{eq:momwvdw} vanishes, and the matrix $K_{{\dot A}B}$
factorizes into a single product of two spinors
\beq
K_{{\dot A}B} = k^\mu\sigma_{\mu,{\dot A}B} = k_{\dot A}k_{B}, \qquad
k_A = \sqrt{2k^0}\, n_{1,A} = \sqrt{2k^0}
\pmatrix{e^{-\ri\phi}\cos\frac{\theta}{2}\cr\sin\frac{\theta}{2}}.
\label{eq:lmomwvdw}
\eeq
In this context, $k_A$ is called a {\it momentum spinor}.

{}Finally, we remark that the decomposition \refeq{eq:momwvdw} is a very
convenient, but not unique, possibility to express a 4-vector $k^\mu$ with
$k^2\ne 0$ in terms
of WvdW spinors. Any splitting of $k^\mu$ into two light-like 4-vectors
yields a decomposition into spinors, since light-like vectors factorize,
as seen above. For instance, choosing an arbitrary light-like
4-vector $a^\mu$ ($a^2=0$) with $a\cdot k\ne 0$, and defining
\beq
\alpha = \frac{k^2}{2a\cdot k}, \qquad
b^\mu = k^\mu -\alpha a^\mu,
\eeq
yields a possible decomposition $k^\mu = \alpha a^\mu + b^\mu$.
In terms of WvdW spinors this corresponds to an arbitrarily chosen
spinor $a_A$ with $K_{\dot CD}a^{\dot C}a^D \ne 0$, leading to the
decomposition
\beq
K_{{\dot A}B} = \alpha\, a_{\dot A}a_B + b_{\dot A}b_B
\qquad \mbox{with} \qquad
b_A = -\frac{K_{{\dot B}A}a^{\dot B}}{\sqrt{K_{{\dot C}D}a^{\dot C}a^D}},
\qquad \alpha = \frac{k^2}{K_{{\dot C}D}a^{\dot C}a^D}.
\label{eq:momwvdw3}
\eeq

\section{Wave functions for helicity eigenstates}
\label{se:wavefunc}

\subsection{\boldmath{Spin-$\frac{1}{2}$ particles}}
\label{spin1/2}

Dirac spinors $\Psi$ belong to the representation 
$D(\frac{1}{2},0)\oplus D(0,\frac{1}{2})$ of the Lorentz group. 
Thus, in terms of WvdW spinors, they are represented by
\beq
\Psi = \pmatrix{\vphantom{\phi_{\dot A}} \phi_A\cr\psi^{\dot A}}.
\label{eq:Psi}
\eeq
The WvdW formalism consistently employs the chiral representation of the
Dirac matrices $\gamma^\mu$:
\beq
\gamma^\mu =
\pmatrix{0&\sigma^\mu_{A{\dot B}}\cr\sigma^{\mu,{\dot A}B}&0}, \qquad
\gamma^5 = \ri\gamma^0\gamma^1\gamma^2\gamma^3 =
\pmatrix{\sigma^0&0\cr 0&-\sigma^0}.
\label{eq:chirdirac}
\eeq
We are interested in plane-wave solutions 
$\Psi=\exp\{\mp\ri kx\}\Psi^{(\pm)}_k$ of
{\it Dirac's equation}
\beq
(\ri\slash\partial-m)\Psi=0,
\label{eq:dirac}
\eeq
which describe the propagation of free Dirac fermions and antifermions,
respectively. Inserting \refeq{eq:Psi} and
\refeq{eq:chirdirac} into Dirac's equation, we obtain the coupled
pair of {\it Weyl equations}
\beq
K_{A{\dot B}}\,\psi^{(\pm),{\dot B}}_k = \pm m \,\phi^{(\pm)}_{k,A},
\qquad
K^{{\dot A}B}\,\phi^{(\pm)}_{k,B} = \pm m \,\psi^{(\pm),{\dot A}}_k,
\qquad
k^2 = m^2.
\label{eq:weyl}
\eeq
Upon applying the decomposition \refeq{eq:momwvdw2} of $K_{{\dot A}B}$,
the following solutions can be easily constructed:
\beq
\Psi^{(\pm)}_{k,1}=\pmatrix{\kappa_{1,A} \cr \mp\kappa_2^{\dot A}}, \qquad
\Psi^{(\pm)}_{k,2}=\pmatrix{\pm\kappa_{2,A} \cr \kappa_1^{\dot A}}.
\label{eq:Psik}
\eeq
The corresponding adjoint spinors $\overline\Psi=\Psi^\dag\gamma_0$ read
\beq
\overline\Psi^{(\pm)}_{k,1}=
\left(\mp\kappa_2^A, \kappa_{1,{\dot A}}\right), \qquad
\overline\Psi^{(\pm)}_{k,2}=
\left(\kappa_1^A, \pm\kappa_{2,{\dot A}}\right).
\label{eq:barPsik}
\eeq
The solutions obey the standard normalization
\beq
\Psi^{(\pm)\dag}_{k,i}\Psi^{(\tau)}_{k,j} =
2k^0\, \delta_{\tau\pm}\, \delta_{ij}, \qquad
\overline\Psi^{(\pm)}_{k,i}\Psi^{(\tau)}_{k,j} =
\pm 2m \delta_{\tau\pm} \delta_{ij}.
\label{eq:norm1/2}
\eeq
Moreover, it is straightforward to check that they indeed form eigenstates 
of the helicity projector
\beq
\Sigma^\pm_k = {\text\frac{1}{2}}(1\pm\gamma^5\slash s_k), \qquad
s^\mu_k = \frac{k^0}{m}\frac{k^\mu}{\vert {\bf k}\vert} -
g^{\mu 0}\frac{m}{\vert {\bf k}\vert}.
\label{eq:spinproj}
\eeq
More precisely, $\Sigma^+_k$ projects onto $\Psi^{(+)}_{k,1}$ and
$\Psi^{(-)}_{k,2}$,  $\Sigma^-_k$ onto $\Psi^{(-)}_{k,1}$
and $\Psi^{(+)}_{k,2}$, i.e.\ $\Psi^{(+)}_{k,1}$ describes a
right-handed fermion, $\Psi^{(-)}_{k,2}$ a right-handed antifermion,
etc.

{}For massless fermions the Weyl equations \refeq{eq:weyl} decouple, and
the helicity eigenstates take the simple form
\beqar
&& \rlap{$ \Psi^{(\pm)}_{k,1} = \pmatrix{k_A\cr 0}, $} \hspace{10em}
\Psi^{(\pm)}_{k,2} = \pmatrix{ 0\cr k^{\dot A}}, \nn\\
&& \rlap{$ \overline\Psi^{(\pm)}_{k,1} = \left(0,k_{\dot A}\right), 
$}\hspace{10em}
\overline\Psi^{(\pm)}_{k,2} = \left(k^A,0\right).
\eeqar

{}For the decomposition \refeq{eq:momwvdw3} of $K_{{\dot A}B}$, also
simple plane-wave solutions exist, namely
\beq
\Psi^{(\pm)}_{k,1} =
\pmatrix{b_A \cr \mp\sqrt{\alpha}\,a^{\dot A}}, \qquad
\Psi^{(\pm)}_{k,2}=\pmatrix{\pm\sqrt{\alpha}\,a_A \cr
b^{\dot A}},
\eeq
which are, however, not related to definite helicity states.

\subsection{Massive spin-1 particles}
\label{spin1}

Spin-1 fields $V^\mu$ transform like ordinary 4-vectors under Lorentz
transformations. This means that polarization vectors $\veps^\mu$
for such fields can be related to 2$\times$2 matrices 
$\veps_{{\dot A}B}$ in the same way as described in \refeq{eq:kmatrix}
for a general 4-vector $k^\mu$. However, one should keep in mind that
polarization vectors need not be real so that $\veps_{{\dot A}B}$ is not
Hermitian in general.

The free field $V^\mu$ of a massive vector particle is governed by
{\it Proca's equation}
\beq
\left[(\partial^2+m^2)g^{\mu\nu}-
\partial^\mu\partial^\nu\right]V_\nu=0,
\eeq
which for $m\ne 0$ is equivalent to the {\it Klein--Gordon equation},
in conjunction with the transversality condition:
\beq
(\partial^2+m^2)V^\mu=0, \qquad \partial\cdot V=0.
\eeq
With the ansatz $V^\mu=\exp\{\mp\ri kx\}\varepsilon_\mu(k)$ 
for particles and antiparticles, respectively, we arrive at
\beq
k^2=m^2, \qquad
k^\nu\varepsilon_\nu(k) =
{\text\frac{1}{2}}K^{{\dot A}B}\varepsilon_{{\dot A}B}(k)=0.
\label{eq:masstrans}
\eeq
{}For $m\ne 0$ there are three linearly independent, space-like polarization
vectors $\varepsilon^\mu_i(k)$, which are usually orthonormalized
according to
\beq
\varepsilon_i^\mu(k)\,\varepsilon_{j,\mu}^*(k)=
{\text\frac{1}{2}} \varepsilon_{i,{\dot A}B}(k)
\varepsilon_j^{* \dot AB}(k) = -\delta_{ij}, \qquad i,j=0,\pm,
\label{eq:epsortho}
\eeq
where $\varepsilon^{* \dot AB}=\varepsilon^*_\mu\sigma^{\mu,\dot AB}$.
A helicity basis for the $\veps^\mu_i(k)$ is given by
\beqar
\varepsilon_{\pm}^\mu(k) & = & \frac{e^{\mp\ri\phi}}{\sqrt{2}}
(0,-\cos\theta\cos\phi\pm\ri\sin\phi,
-\cos\theta\sin\phi\mp\ri\cos\phi,\sin\theta), \nn \\
\varepsilon_0^\mu(k) & = & s^\mu_k =
\frac{k^0}{m}\biggl(\frac{\vert {\bf k}\vert}{k^0},
\cos\phi\sin\theta,\sin\phi\sin\theta,\cos\theta\biggr),
\label{eq:polvectors}
\eeqar
where $s^\mu_k$ is related to $k^\mu$ as given in \refeq{eq:spinproj}.
In terms of WvdW spinors this spin basis reads
\beqar
\veps_{+,{\dot A}B}(k) & = & \sqrt{2}\,n_{2,{\dot A}}\,n_{1,B}, \qquad
\veps_{-,{\dot A}B}(k)   =   \sqrt{2}\,n_{1,{\dot A}}\,n_{2,B}, \nn \\
\veps_{0,{\dot A}B}(k) & = & \frac{1}{m}
(\kappa_{1,{\dot A}}\,\kappa_{1,B}-\kappa_{2,{\dot A}}\,\kappa_{2,B}),
\label{eq:polspinors}
\eeqar
where the spinors $n_{i,A}$ and $\kappa_{i,A}$ are defined in 
\refeq{eq:momwvdw} and \refeq{eq:momwvdw2}, respectively.
Note that care has to be taken when dealing with conjugate polarization
vectors, which, in particular, occur for outgoing spin-1 particles in
transition amplitudes. In such cases the transition from the 4-vector 
$\veps^{*\mu}$ to the 2$\times$2 matrix $\veps^*_{{\dot A}B}$ upon
contraction with the $\sigma$ matrices has to be consistently performed 
for the conjugated polarization vector, i.e.\
$\veps^*_{{\dot A}B} = \veps^*_\mu\sigma^\mu_{\dot AB}$,
which in general is different from taking the complex conjugate of
$\veps_{{\dot A}B}$. For clarity, we give also the helicity basis for 
an outgoing spin-1 particle:
\beqar
\veps^*_{+,{\dot A}B}(k) & = & \sqrt{2}\,n_{1,{\dot A}}\,n_{2,B}, \qquad
\veps^*_{-,{\dot A}B}(k)   =   \sqrt{2}\,n_{2,{\dot A}}\,n_{1,B}, \nn \\
\veps^*_{0,{\dot A}B}(k) & = & \frac{1}{m}
(\kappa_{1,{\dot A}}\,\kappa_{1,B}-\kappa_{2,{\dot A}}\,\kappa_{2,B}),
\label{eq:cpolspinors}
\eeqar
which correspond to the conjugate polarization vectors $\veps^*_i$ of
$\veps_i$ given in \refeq{eq:polvectors}. From \refeq{eq:polspinors}
and \refeq{eq:cpolspinors} we obtain the relations
\beq
\veps_{i,{\dot A}B}(k) = \veps_{-i,B{\dot A}}(k) = 
\veps^*_{-i,{\dot A}B}(k) = \veps^*_{i,B{\dot A}}(k).
\eeq

{}Finally, we give a simple spin basis for the decomposition 
\refeq{eq:momwvdw3} of $K_{{\dot A}B}$,
\beq
\varepsilon_{+,{\dot A}B}(p) =
\frac{\sqrt{2}\,a_{\dot A}\,b_B}{\langle ab\rangle^*}, \quad
\varepsilon_{-,{\dot A}B}(p)=
\frac{\sqrt{2}\,b_{\dot A}\,a_B}{\langle ab\rangle}, \quad
\varepsilon_{0,{\dot A}B}(p) = \frac{1}{m}
(b_{\dot A}\,b_B-\alpha\,a_{\dot A}a_B),
\eeq
which is, however, not related to helicity eigenstates.

\subsection{Massless spin-1 particles}

In contrast with the case of spin-$\frac{1}{2}$ particles, the zero-mass
limit for vector bosons confronts us with a new physical situation. For 
$m\to 0$ the polarization vector $\veps^\mu_0$ does not exist, since the 
longitudinal polarization does not yield a physical state for massless 
spin-1 particles. The corresponding degree of freedom manifests itself 
in the arbitrariness of gauge for the two polarization vectors 
$\veps^\mu_{\pm}$. In the language of WvdW spinors this fact is expressed 
very elegantly:
\beqar
\veps_{+,{\dot A}B}(k) & = & 
\frac{\sqrt{2}\,g_{+,{\dot A}}\,k_B}{\langle g_+k\rangle^*}, \qquad
\veps_{-,{\dot A}B}(k)   =   
\frac{\sqrt{2}\,k_{\dot A}\,g_{-,B}}{\langle g_-k\rangle},
\nn\\
\veps^*_{+,{\dot A}B}(k) & = & 
\frac{\sqrt{2}\,k_{\dot A}\,g_{+,B}}{\langle g_+k\rangle}, \qquad
\veps^*_{-,{\dot A}B}(k)   =   
\frac{\sqrt{2}\,g_{-,{\dot A}}\,k_B}{\langle g_-k\rangle^*}, 
\label{eq:m0polspinors}
\eeqar
where $g_{\pm,A}$ denote arbitrary spinors with 
$\langle g_\pm k\rangle\ne 0$; they are called {\it gauge spinors}.
The difference of two matrices $\veps_{i,\dot AB}(k)$ for differently 
chosen gauge spinors is proportional to the momentum matrix 
$k_{\dot A}k_B$, as can be easily checked by applying the Schouten 
identity \refeq{eq:schouten}. The transversality condition
$k\cdot\veps_\pm=0$ is still fulfilled for any $g_{\pm,A}$,
i.e.\ the Lorentz gauge is maintained for the $\veps_\pm$ of 
(\ref{eq:m0polspinors}). 
Transition amplitudes do not depend on the choice of gauge
spinors. This means that the $g_{\pm,A}$ can be appropriately chosen 
to simplify the algebraic expression. Note that it is even
possible to take different sets of gauge spinors for each helicity
channel.

\section{Calculation of helicity amplitudes}
\label{se:helamp}

\subsection{Feynman rules}

{}Feynman rules are usually written down in terms of 4-vectors and Dirac
matrices. Here we describe how such Feynman rules can be directly 
translated into the language of WvdW spinors. Obviously this procedure
is much more practical than going back to the underlying Lagrangian and
introducing WvdW spinors there.

\begin{table}
\setlength{\unitlength}{1pt}
\begin{center}
\begin{tabular}{|l|l|}
\hline
Incoming fields: & Outgoing fields:
\\
\hline
\begin{picture}(190,25)
\ArrowLine(  5,10)(45,10)
\Vertex   ( 45,10){2.0}
\Text     (110,10)[l]{$\Psi^{(+)}_{k,i}$}
\end{picture}
&
\begin{picture}(190,25)
\ArrowLine(  5,10)(45,10)
\Vertex   (  5,10){2.0}
\Text     (110,10)[l]{$\overline\Psi^{(+)}_{k,i}$}
\end{picture}
\\
\hline
\begin{picture}(190,25)
\ArrowLine( 45,10)(  5,10)
\Vertex   ( 45,10){2.0}
\Text     (110,10)[l]{$\overline\Psi^{(-)}_{k,i}$}
\end{picture}
&
\begin{picture}(190,25)
\ArrowLine( 45,10)(  5,10)
\Vertex   (  5,10){2.0}
\Text     (110,10)[l]{$\Psi^{(-)}_{k,i}$}
\end{picture}
\\
\hline
\begin{picture}(190,25)
\Photon( 45,10)(  5,10){1.5}{5}
\Vertex   ( 45,10){2.0}
\Text     (  9,18)[l]{$_{AB}$}
\Text     ( 10, 3)[l]{$_\mu$}
\Text     ( 65,10)[l]{$\veps_{i,\mu}(k)$}
\Text     (110,10)[l]{$\longrightarrow$}
\Text     (140,10)[l]{$\veps_{i,{\dot A}B}(k)$}
\end{picture}
&
\begin{picture}(190,25)
\Photon( 45,10)(  5,10){1.5}{5}
\Vertex   (  5,10){2.0}
\Text     ( 32,18)[l]{$_{AB}$}
\Text     ( 35, 3)[l]{$_\mu$}
\Text     ( 65,10)[l]{$\veps^*_{i,\mu}(k)$}
\Text     (110,10)[l]{$\longrightarrow$}
\Text     (140,10)[l]{$\veps^*_{i,{\dot A}B}(k)$}
\end{picture}
\\
\hline\hline
\multicolumn{2}{|l|}{ 
Propagators: }
\\
\hline
\multicolumn{2}{|l|}{ 
\begin{picture}(380,35)
\ArrowLine(  5,15)( 45,15)
\Vertex   (  5,15){2.0}
\Vertex   ( 45,15){2.0}
\Text     (  5,23)[l]{$_B$}
\Text     ( 23,24)[l]{$k$}
\Text     ( 40,23)[l]{$_A$}
\Text     ( 65,15)[l]{$\disp\frac{\ri(\slash k+m_f)}{k^2-m_f^2}$}
\Text     (140,15)[l]{$\longrightarrow$}
\Text     (180,15)[l]{$\disp\frac{\ri}{k^2-m_f^2}
		\pmatrix{m_f\delta_A^B & K_{A{\dot B}} \\
		K^{{\dot A}B} & m_f\delta_{\dot B}^{\dot A}}$ }
\end{picture} }
\\
\hline
\multicolumn{2}{|l|}{ 
\begin{picture}(380,35)
\Photon   (  5,15)( 45,15){1.5}{5}
\Vertex   (  5,15){2.0}
\Vertex   ( 45,15){2.0}
\Text     (  5,23)[l]{$_{AB}$}
\Text     ( 35,23)[l]{$_{CD}$}
\Text     (  5, 8)[l]{$_\mu$}
\Text     ( 40, 8)[l]{$_\nu$}
\Text     ( 65,15)[l]{$\disp\frac{-\ri g_{\mu\nu}}{k^2-M_V^2}$}
\Text     (140,15)[l]{$\longrightarrow$}
\Text     (180,15)[l]{$\disp\frac{-2\ri\epsilon_{{\dot A}{\dot C}}\epsilon_{BD}}
		{k^2-M_V^2}$ }
\end{picture} }
\\
\hline
\end{tabular}
\end{center}
\caption{Prescriptions for translating ordinary Feynman rules for
external fields and propagators into the WvdW formalism.}
\label{ta:exproprul} 
\end{table}
\begin{table}
\setlength{\unitlength}{1pt}
\begin{center}
\begin{tabular}{|l|}
\hline
Vertices: \\
\hline
\begin{picture}(430,55)
\ArrowLine(  5,45)( 45,25)
\ArrowLine( 45,25)(  5, 5)
\Photon   ( 45,25)( 85,25){1.5}{5}
\Vertex   ( 45,25){2.0}
\Text     ( 75,34)[l]{$_{CD}$}
\Text     ( 75,16)[l]{$_\mu$}
\Text     (  5,36)[l]{$_B$}
\Text     (  5,15)[l]{$_A$}
\Text     ( 25, 8)[l]{$_1$}
\Text     ( 25,42)[l]{$_2$}
\Text     (110,25)[l]{$\disp\ri e\gamma^\mu C^\tau_{V\bar f_1f_2}\omega_\tau $}
\Text     (220,25)[l]{$\longrightarrow$}
\Text     (260,25)[l]{$\disp\ri e
	\pmatrix{0 & C^-_{V\bar f_1f_2}\delta^{\dot C}_{\dot B}\delta^D_A \\
	C^+_{V\bar f_1f_2}\epsilon^{{\dot A}{\dot C}}\epsilon^{BD} & 0}$} 
\end{picture}
\\
\hline
\begin{picture}(430,55)
\ArrowLine(  5,45)( 45,25)
\ArrowLine( 45,25)(  5, 5)
\DashLine ( 45,25)( 85,25){5}
\Vertex   ( 45,25){2.0}
\Text     (  5,36)[l]{$_B$}
\Text     (  5,15)[l]{$_A$}
\Text     ( 25, 8)[l]{$_1$}
\Text     ( 25,42)[l]{$_2$}
\Text     (110,25)[l]{$\disp\ri e C^\tau_{S\bar f_1f_2}\omega_\tau $}
\Text     (220,25)[l]{$\longrightarrow$}
\Text     (260,25)[l]{$\disp\ri e
	\pmatrix{C^+_{S\bar f_1f_2}\delta_A^B & 0 \\
		0 & C^-_{S\bar f_1f_2}\delta_{\dot B}^{\dot A}}$} 
\end{picture}
\\
\hline
\begin{picture}(430,65)
\Photon   (  5,50)( 45,30){1.5}{5}
\Photon   ( 45,30)(  5,10){1.5}{5}
\Photon   ( 45,30)( 85,30){1.5}{5}
\Vertex   ( 45,30){2.0}
\Text     ( 75,39)[l]{$_{EF}$}
\Text     ( 75,21)[l]{$_\rho$}
\Text     (  3,40)[l]{$_{AB}$}
\Text     (  9,55)[l]{$_\mu$}
\Text     (  3,21)[l]{$_{CD}$}
\Text     (  9, 6)[l]{$_\nu$}
\Text     ( 28,15)[l]{$_2$}
\Text     ( 28,47)[l]{$_1$}
\Text     ( 60,38)[l]{$_3$}
\Text     (110,50)[l]{$\ri eC_{V_1V_2V_3} \left[g^{\mu\nu}(k_1-k_2)^\rho
		+g^{\nu\rho}(k_2-k_3)^\mu+g^{\mu\rho}(k_3-k_1)^\nu\right] $}
\Text     (110,30)[l]{$\longrightarrow$}
\Text     (150,30)[l]{$\frac{\ri}{4}eC_{V_1V_2V_3}
	\left[\epsilon^{{\dot A}{\dot C}}\epsilon^{BD}(K_1-K_2)^{{\dot E}F}
     	+\epsilon^{{\dot C}{\dot E}}\epsilon^{DF}(K_2-K_3)^{{\dot A}B}\right.$}
\Text     (202,10)[l]{$\left.{}
     	+\epsilon^{{\dot A}{\dot E}}\epsilon^{BF}(K_3-K_1)^{{\dot C}D}\right]$}
\end{picture}
\\
\hline
\begin{picture}(430,65)
\Photon   (  5,50)( 45,30){1.5}{5}
\Photon   ( 45,30)(  5,10){1.5}{5}
\DashLine ( 45,30)( 85,30){5}
\Vertex   ( 45,30){2.0}
\Text     (  3,40)[l]{$_{AB}$}
\Text     (  9,55)[l]{$_\mu$}
\Text     (  3,21)[l]{$_{CD}$}
\Text     (  9, 6)[l]{$_\nu$}
\Text     ( 28,15)[l]{$_2$}
\Text     ( 28,47)[l]{$_1$}
\Text     (110,30)[l]{$\ri eC_{SV_1V_2} g^{\mu\nu}$}
\Text     (220,30)[l]{$\longrightarrow$}
\Text     (260,30)[l]{$\frac{\ri}{2}eC_{SV_1V_2}
		\epsilon^{{\dot A}{\dot C}}\epsilon^{BD}$}
\end{picture}
\\
\hline
\begin{picture}(430,65)
\DashLine (  5,50)( 45,30){5}
\DashLine ( 45,30)(  5,10){5}
\Photon   ( 45,30)( 85,30){1.5}{5}
\Vertex   ( 45,30){2.0}
\Text     ( 75,39)[l]{$_{AB}$}
\Text     ( 75,21)[l]{$_\mu$}
\Text     ( 28,15)[l]{$_2$}
\Text     ( 28,47)[l]{$_1$}
\Text     (110,30)[l]{$\ri eC_{VS_1S_2}(k_1-k_2)^\mu$}
\Text     (220,30)[l]{$\longrightarrow$}
\Text     (260,30)[l]{$\frac{\ri}{2}eC_{VS_1S_2}(K_1-K_2)^{{\dot A}B}$}
\end{picture}
\\
\hline
\begin{picture}(430,65)
\Photon   (  5,50)( 45,30){1.5}{5}
\Photon   ( 45,30)(  5,10){1.5}{5}
\Photon   ( 45,30)( 85,50){1.5}{5}
\Photon   ( 45,30)( 85,10){1.5}{5}
\Vertex   ( 45,30){2.0}
\Text     ( 75,40)[l]{$_{GH}$}
\Text     ( 75,55)[l]{$_\sigma$}
\Text     (  3,40)[l]{$_{AB}$}
\Text     (  9,55)[l]{$_\mu$}
\Text     (  3,21)[l]{$_{CD}$}
\Text     (  9, 6)[l]{$_\nu$}
\Text     ( 75,21)[l]{$_{EF}$}
\Text     ( 75, 6)[l]{$_\rho$}
\Text     ( 60,15)[l]{$_3$}
\Text     ( 28,15)[l]{$_2$}
\Text     ( 28,47)[l]{$_1$}
\Text     ( 60,47)[l]{$_4$}
\Text     (110,50)[l]{$\ri e^2C_{V_1V_2V_3V_4} \left[2g^{\mu\nu}g^{\rho\sigma}
	-g^{\mu\rho}g^{\nu\sigma}-g^{\mu\sigma}g^{\nu\rho} \right] $}
\Text     (110,30)[l]{$\longrightarrow$}
\Text     (150,30)[l]{$\frac{\ri}{4}e^2C_{V_1V_2V_3V_4} \left[
	2\epsilon^{{\dot A}{\dot C}}\epsilon^{BD}
	\epsilon^{{\dot E}{\dot G}}\epsilon^{FH} 
	-\epsilon^{{\dot A}{\dot E}}\epsilon^{BF}
	\epsilon^{{\dot C}{\dot G}}\epsilon^{DH} \right.$}
\Text     (202,10)[l]{$\left.{}
	-\epsilon^{{\dot A}{\dot G}}\epsilon^{BH}
	\epsilon^{{\dot C}{\dot E}}\epsilon^{DF} \right]$}
\end{picture}
\\
\hline
\begin{picture}(430,65)
\Photon   (  5,50)( 45,30){1.5}{5}
\Photon   ( 45,30)(  5,10){1.5}{5}
\DashLine ( 45,30)( 85,50){5}
\DashLine ( 45,30)( 85,10){5}
\Vertex   ( 45,30){2.0}
\Text     (  3,40)[l]{$_{AB}$}
\Text     (  9,55)[l]{$_\mu$}
\Text     (  3,21)[l]{$_{CD}$}
\Text     (  9, 6)[l]{$_\nu$}
\Text     ( 60,15)[l]{$_3$}
\Text     ( 28,15)[l]{$_2$}
\Text     ( 28,47)[l]{$_1$}
\Text     ( 60,47)[l]{$_4$}
\Text     (110,30)[l]{$\ri e^2C_{V_1V_2S_3S_4} g^{\mu\nu}$}
\Text     (220,30)[l]{$\longrightarrow$}
\Text     (260,30)[l]{$\frac{\ri}{2}e^2C_{V_1V_2S_3S_4}
		\epsilon^{{\dot A}{\dot C}}\epsilon^{BD}$}
\end{picture}
\\
\hline
\end{tabular}
\end{center}
\caption{Prescriptions for translating ordinary Feynman rules for
vertices into the WvdW formalism. Scalar fields, fermion fields, and vector
fields are generically denoted by $S$, $f$, and $V$, respectively, and
$\omega_\pm=\frac{1}{2}(1\pm\gamma^5)$ are the chirality projectors.}
\label{ta:vertrul}
\end{table}
Given any Feynman graph, the trick is to contract each
vector-boson leg of a vertex with the identity
$\delta^\mu_\nu = \frac{1}{2}\sigma_\nu^{{\dot A}B}\sigma^{\mu}_{{\dot A}B}$
and to shift the factor $\sigma^{\mu}_{{\dot A}B}$ to the vector-boson
propagator or to the external wave function of the vector boson that is 
attached to this vertex. Moreover, for each occurring Dirac matrix the chiral
representation \refeq{eq:chirdirac} has to be used. 
In particular, the unit matrix ${\bf 1}$ in the Dirac space and a slashed 
quantity $\as$ read
\beq
{\bf 1} = \pmatrix{\delta_A^B & 0 \cr 0 & \delta_{\dot B}^{\dot A}},
\qquad
\as = \pmatrix{0 & a_{A{\dot B}} \cr a^{{\dot A}B} & 0 },
\eeq
respectively.
Tables~\ref{ta:exproprul} and \ref{ta:vertrul} summarize the necessary
changes of the generic Feynman rules for the electroweak Standard Model
in the 't~Hooft--Feynman gauge. The explicit values for the couplings can,
for instance, be found in \citere{sm}. The new Feynman rules for other
conventions or different models can easily be worked out by the reader.

Once the Feynman rules are settled, it is very easy to explicitly write
down helicity amplitudes for any process at tree level. Expressing all
particle momenta in terms of spinors, as described in \refse{se:4vectors},
each helicity amplitude is reduced to an algebraic expression in terms
of antisymmetric spinor products $\langle\phi\psi\rangle$ after all
spinor indices are contracted. These spinor contractions can be
performed like usual contractions of Lorentz indices, apart from taking
care of the antisymmetry. The form of the amplitudes obtained this
way is already well suited to numerical evaluations.

\subsection{Discrete symmetries}
\label{se:discsym}

Discrete symmetries relate helicity amplitudes of 
different processes or of one and the same process. Thus, they
either provide convenient cross checks for results or they allow for a
reduction of the algebraic work by generating various amplitudes from a
generic set of amplitudes. In the following we show how to derive the
relations implied by crossing symmetry, parity, and CP symmetry, which 
are the most important discrete symmetries in practice. 
\renewcommand{\labelenumi}{(\roman{enumi})}
\begin{enumerate}
\item Crossing symmetry: \\
Crossing symmetry transforms an incoming particle into the corresponding
outgoing antiparticle, or vice versa. Denoting the momentum of a given
particle by $k$, the inversion $k\to -k$ can be consistently obtained by
substituting 
\beq
\kappa_{i,\dot A}\to -\kappa_{i,\dot A}, 
\qquad
\kappa_{i,A}\to +\kappa_{i,A}
\label{eq:crossing}
\eeq
in the decomposition \refeq{eq:momwvdw2} of $k$, i.e.\ by inverting the
contravariant parts only. In order to relate helicity amplitudes, it is
also necessary to consider the relation between the wave functions of
the respective incoming and outgoing fields. Inspecting the explicit
form of the wave functions for the helicity eigenstates of 
\refse{se:wavefunc}, one finds that the substitution \refeq{eq:crossing}
transforms incoming (outgoing) particles 
into outgoing (incoming) antiparticles with reversed
helicity modulo sign change.
Specifically, there is a global factor $-1$ for each crossed spin-1 field 
and a factor $\pm{\mathrm{sgn}}(\sigma)$ for each incoming/outgoing fermion
with helicity $\sigma$ that is involved in the crossing.
\item Parity: \\
If parity is a symmetry, every helicity amplitude, up to a
phase factor, agrees with the corresponding amplitude with opposite
helicities, after the spatial parts of all momenta are inverted. 
At the level of WvdW spinors, the inversion of the spatial parts of 
momenta is connected with the interchange of the spinors $\psi_A$ and
$\psi^{\dot A}$. It turns out that
each helicity amplitude, up to a global sign factor, agrees with the
complex conjugate amplitude with opposite helicity configuration.
More precisely, if $\ri\M(\sigma_i;\lambda_j)$ is an amplitude involving 
$n$ fermions and $\bar n$ antifermions with helicities $\sigma_i$, and
$n_V$ vector bosons with helicities $\lambda_j$, we get the relations
\beq
\M(-\sigma_i;-\lambda_j) = 
(-1)^{\bar n}\sgn(\sigma_1\cdots\sigma_{n+\bar n}) \M(\sigma_i;\lambda_j)^*
\label{eq:MP}
\eeq
for tree-level amplitudes.
If parity is violated, these relations for amplitudes remain valid if 
left- and right-handed couplings are appropriately substituted. The
explicit derivation of \refeq{eq:MP} and the modifications for broken
parity can be found in the appendix.
\item CP symmetry: \\
CP is the product of parity, which is explained above, and charge
conjugation, which interchanges particles with the respective
antiparticles. Therefore, CP symmetry leads to relations between the
helicity amplitudes of a given process and the complex conjugate
helicity amplitudes with reversed helicities of the process involving
the respective antiparticles. We consider an amplitude 
$\ri\M(\sigma_i;\lambda_j)$ and the one for the CP-related process,
$\ri\overline{\M}(\sigma_i;\lambda_j)$, 
where the respective helicities for the fermions and vector bosons are
assigned as above.
If CP is an exact symmetry these two matrix elements are related by
\beq
\overline{\M}(-\sigma_i;-\lambda_j) = 
(-1)^{n_V+n_{\mathrm{in}}+\bar n_{\mathrm{in}}} 
\sgn(\sigma_1\cdots\sigma_{n+\bar n}) \M(\sigma_i;\lambda_j)^*
\label{eq:MCP}
\eeq
at tree level,
where $n_{\mathrm{in}}+\bar n_{\mathrm{in}}$ is the total number of incoming
fermions and antifermions. The explicit derivation of these relations
and their modifications if CP is violated are given in the appendix.

\end{enumerate}
The actual use of these symmetries will be illustrated when we consider
helicity amplitudes of concrete processes, in \refse{se:app}.

\subsection{Remarks on the choice of gauge spinors}
\label{se:gspinor}

Since a helicity amplitude does not depend on the actual insertions for 
gauge spinors, they can, in principle, be chosen arbitrarily. Usually a gauge
spinor $g_A$ is chosen such that as many terms as possible in an 
amplitude vanish. However, as can be seen from \refeq{eq:m0polspinors}, an
arbitrary choice of $g_A$ in general leads to unphysical poles
at the zeros of $\langle gk\rangle$, in
individual terms that contribute to an amplitude. Of course, the
unphysical pole drops out in the final result, but this cancellation can
cause instabilities in a numerical evaluation. 

Unphysical poles can, for instance,
consistently be avoided by setting $g_A=n_{2,A}$, where
$n_{2,A}$ is related to the momentum $k$ of the massless spin-1 particle
as specified in \refeq{eq:momwvdw}. This choice is identical with 
\refeq{eq:polspinors} for the transverse modes of a massive spin-1 particle.
The drawback of this choice is that no algebraic simplifications result.

Another possibility to avoid numerical problems that are due to
unphysical poles is to cancel such poles analytically, before the
numerical evaluation. In general this task can be extremely cumbersome, 
and the additional work devalues preceding simplifications. In the
following we give a very convenient choice of $g_A$, yielding desirable
simplifications without leaving uncancelled unphysical poles.

\begin{figure}
\setlength{\unitlength}{1pt}
\begin{center}
\begin{picture}(110,70)
\ArrowLine(50,50)(90,50)
\ArrowLine(50,10)(50,50)
\Photon   (50,50)(10,50){2}{4}
\Vertex   (50,50){2.0}
\Vertex   (50,10){2.0}
\Text     (10, 0)[l]{(a)}
\LongArrow(22,58)(38,58)
\LongArrow(62,58)(78,58)
\LongArrow(58,22)(58,38)
\Text     (26,66)[l]{$k$}
\Text     (66,66)[l]{$p$}
\Text     (62,30)[l]{$p-k$}
\Text     (10,42)[l]{$_{AB}$}
\Text     (82,44)[l]{$_C$}
\Text     (40,42)[l]{$_D$}
\Text     (40,17)[l]{$_E$}
\end{picture}
\begin{picture}(110,70)
\ArrowLine(90,50)(50,50)
\ArrowLine(50,50)(50,10)
\Photon   (50,50)(10,50){2}{4}
\Vertex   (50,50){2.0}
\Vertex   (50,10){2.0}
\Text     (10, 0)[l]{(b)}
\LongArrow(22,58)(38,58)
\LongArrow(62,58)(78,58)
\LongArrow(58,22)(58,38)
\Text     (26,66)[l]{$k$}
\Text     (66,66)[l]{$p$}
\Text     (62,30)[l]{$p-k$}
\Text     (10,42)[l]{$_{AB}$}
\Text     (82,44)[l]{$_C$}
\Text     (40,42)[l]{$_D$}
\Text     (40,17)[l]{$_E$}
\end{picture}
%
\end{center}
\caption{Subdiagrams containing the fermion propagator denominator
\mbox{$[(p-k)^2-\Mf^2]^{-1}$} for processes with an incoming photon of
momentum $k$ ($k^2=0$) and an outgoing fermion or antifermion of
momentum $p$ ($p^2=\Mf^2$).}
\label{fi:aff}
\end{figure}
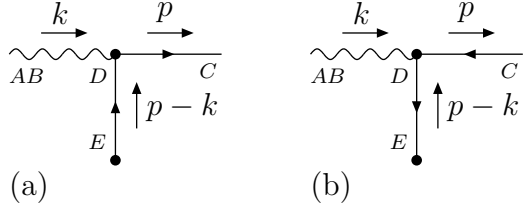

Consider a process with an incoming photon of momentum $k$ and an
outgoing fermion $\Pf$ of momentum $p$; the momentum matrix for $p$ is
denoted by
\beq
P_{\dot AB} = \sum_{i=1,2} \kappa_{i,\dot A} \kappa_{i,B}.
\label{eq:pdecomp}
\eeq
We choose the same gauge spinor
\beq
g_A = P_{A\dot B} k^{\dot B} = 
\sum_{i=1,2} \kappa_{i,A} \langle\kappa_i k\rangle^*
\label{eq:gsp}
\eeq
for both photon helicities. The problematic denominators contained in the
photon polarization vectors \refeq{eq:m0polspinors} are given by
\newcommand{\MPK}{(p\cdot k)}
\beq
\langle gk \rangle = k^A P_{A\dot B} k^{\dot B} = 2\MPK.
\eeq
This has the form of the inverse propagator $(p-k)^2-\Mf^2=-2\MPK$
that appears if the incoming photon directly couples to the outgoing
fermion $\Pf$ so that no unphysical pole is introduced at all. 
The subdiagram containing this propagator is shown in \reffi{fi:aff}a.
In Feynman graphs that contain this subdiagram, we obtain the factor
$\MPK^{-2}$, one power is due to the propagator, another is due to
the polarization vector.
{}For small $\MPK$ we again get an unwanted cancellation of one power in 
$\MPK$. However, this cancellation can be easily performed analytically. 
Denoting the Dirac spinor of the outgoing fermion with helicity
$\sigma$, generically, by
\beq
\overline\Psi^{(+)}_{p,i} = (\psi^C,\phi_{\dot C})
\qquad \mbox{with} \qquad 
(\phi,\psi) = \Biggl\{
\barr{ll} (\kappa_1,-\kappa_2) & \quad \mbox{for} \; i=1 \; (\sigma=+), \\
          (\kappa_2, \kappa_1) & \quad \mbox{for} \; i=2 \; (\sigma=-), \earr
\label{eq:Psif}
\eeq
and using the wave function $\veps_{\dot AB}$ for the photon with
helicity $\lambda$, the subdiagram reads
\beq
T^{\gamma\Pf}_\lambda(\sigma) = 
-\frac{Q_\Pf e}{2\MPK}
\left( \psi^C,\phi_{\dot C} \right) 
\pmatrix{ 0 & \delta^{\dot A}_{\dot D}\delta^B_C \cr
	  \epsilon^{\dot A\dot C}\epsilon^{BD} & 0 }
\veps_{\dot AB} 
\pmatrix{ \Mf\delta^E_D & (p-k)_{D\dot E} \cr
	  (p-k)^{\dot DE} & \Mf\delta^{\dot D}_{\dot E} },
\eeq
where $Q_f$ is the relative electromagnetic charge of the fermion $f$.
Upon inserting \refeq{eq:pdecomp},
\refeq{eq:gsp}, and the explicit expressions \refeq{eq:m0polspinors} for 
$\veps_{+,\dot AB}$, we get
\beq
T^{\gamma\Pf}_+(\sigma) = \frac{Q_\Pf e}{2\sqrt{2}\MPK^2} \langle k\psi\rangle
\left( 2\MPK k^E +
\Mf\left[\langle\phi\psi\rangle k^E-\langle k\psi\rangle\phi^E
+\langle k\phi\rangle\psi^E\right], 0 \right).
\label{eq:tp} 
\eeq
Here we have used that $P_{\dot AB}=\phi_{\dot A}\phi_B+\psi_{\dot A}\psi_B$
and $\langle\phi\psi\rangle=\Mf$ for both fermion helicities. 
The expression within square brackets in \refeq{eq:tp} 
vanishes according to Schouten's identity \refeq{eq:schouten}
so that one factor $\MPK$ cancels.
Performing a similar calculation for $T^{\gamma\Pf}_-(\sigma)$, 
we find the simple results
\beq
T^{\gamma\Pf}_+(\sigma) = \frac{Q_\Pf e}{\sqrt{2}\MPK} \langle k\psi\rangle
\Big( k^E, 0\Big), 
\qquad
T^{\gamma\Pf}_-(\sigma) = \frac{Q_\Pf e}{\sqrt{2}\MPK} \langle k\phi\rangle^*
\Big( 0, k_{\dot E}\Big). 
\label{eq:Taf}
\eeq
Considering an outgoing antifermion instead of a fermion, 
as shown in \reffi{fi:aff}b, we get 
\beq
T^{\gamma\bar\Pf}_+(\sigma) = -\frac{Q_\Pf e}{\sqrt{2}\MPK} 
\langle k\phi\rangle \pmatrix{ k_E \cr 0 },
\qquad
T^{\gamma\bar\Pf}_-(\sigma) = -\frac{Q_\Pf e}{\sqrt{2}\MPK} 
\langle k\psi\rangle^* \pmatrix{ 0 \cr k^{\dot E} },
\label{eq:Tafbar}
\eeq
where the Dirac spinor for the antifermion is generically denoted by
\beq
\Psi^{(-)}_{p,i} = 
\pmatrix{ \vphantom{\phi_{\dot C}}\phi_C \cr \psi^{\dot C} }
\qquad \mbox{with} \qquad 
(\phi,\psi) = \Biggl\{
\barr{ll} ( \kappa_1,\kappa_2) & \quad \mbox{for} \; i=1 \; (\sigma=-), \\
          (-\kappa_2,\kappa_1) & \quad \mbox{for} \; i=2 \; (\sigma=+). \earr
\label{eq:Psifbar}
\eeq
{}From the results \refeq{eq:Taf} and \refeq{eq:Tafbar} for the
subdiagrams (a) and (b) of \reffis{fi:aff}, respectively, all similar
subdiagrams involving outgoing photons and/or incoming (anti)fermions
follow by crossing, as described in the previous section. 

The above procedure also works if the (anti)fermions in \reffi{fi:aff}
are replaced by charged gauge bosons, e.g.\ by the W~boson of the 
Electroweak Standard Model. In this case, however, one has to take into
account the contribution of the associated would-be Goldstone boson,
which appears on internal lines (vertical lines in \reffi{fi:aff}).

\section{Applications}
\label{se:app}

In order to illustrate the actual use of the formalism described above, 
we calculate full sets of helicity amplitudes for some processes with 
massive particles and photons in lowest order. The presented results have 
been derived analytically and simplified as far as possible; this 
demonstrates the analytical power of the technique. In addition, the
amplitudes have been evaluated by performing the spinor contractions 
automatically in {\sl Mathematica} \cite{math}, which is very simple.
The results of both approaches have been compared numerically.

The following examples only include massive external fermions. For an
application of the helicity basis for massive spin-1 particles we refer to
\citere{bo93}, where the radiative processes 
$\Pem\gamma\to\PWm\nu_\Pe\gamma, \Pem\PZ\gamma$ are discussed.

\subsection{The process \boldmath{$\gamma\gamma\to f\bar f$}}

\begin{figure}
\setlength{\unitlength}{1pt}
\begin{center}
\begin{picture}(110,70)
\ArrowLine(50,50)(90,50)
\ArrowLine(50,10)(50,50)
\ArrowLine(90,10)(50,10)
\Photon   (50,50)(10,50){2}{4}
\Photon   (50,10)(10,10){2}{4}
\Vertex   (50,50){2.0}
\Vertex   (50,10){2.0}
\Text     (10,-5)[l]{(a)}
\end{picture}
\begin{picture}(110,70)
\ArrowLine(50,50)(90,50)
\ArrowLine(50,10)(50,50)
\ArrowLine(90,10)(50,10)
\Photon   (50,50)(10,10){2}{6}
\Photon   (50,10)(10,50){2}{6}
\Vertex   (50,50){2.0}
\Vertex   (50,10){2.0}
\Text     (10,-5)[l]{(b)}
\end{picture}
\end{center}
\caption{Born diagrams for $\gamma\gamma\to f\bar f$.}
\label{fi:aaff}
\end{figure}
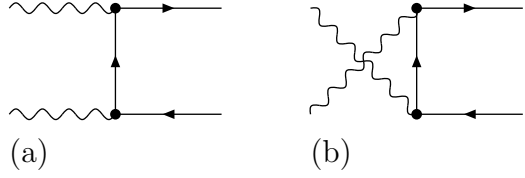

{}For illustration we start with the simple QED process
\beq
\gamma(k_1,\lambda_1) + \gamma(k_2,\lambda_2) \; \longrightarrow \;
f(p,\sigma) + \bar f(p',\sigma').
\eeq
The momentum matrix and the wave function for the outgoing fermion $\Pf$
are denoted as in \refeq{eq:pdecomp} and \refeq{eq:Psif}, respectively.
{}For the outgoing antifermion we define
\beq
P'_{\dot AB} = \sum_{i=1,2} \kappa'_{i,\dot A} \kappa'_{i,B}
\label{eq:ppdecomp}
\eeq
and
\beq
\Psi^{(-)}_{p'} = 
\pmatrix{ \vphantom{\phi_{\dot C}}\phi'_C \cr \psi^{\prime\dot C} }
\qquad \mbox{with} \qquad 
(\phi',\psi') = \Biggl\{
\barr{ll} ( \kappa'_1,\kappa'_2) & \quad \mbox{for} \; \sigma'=-, \\
          (-\kappa'_2,\kappa'_1) & \quad \mbox{for} \; \sigma'=+. \earr
\label{eq:Psifbar2}
\eeq
Since we want to make use of the results of the previous section, we set
the gauge spinor $g_{n,A}$ for the $n$th photon to
\beq
g_{n,A} = P_{A\dot B} k^{\dot B}_n = 
\sum_{i=1,2} \kappa_{i,A} \langle\kappa_i k_n\rangle^*,
\label{eq:gsp12}
\qquad n=1,2.
\eeq
In lowest order the two Feynman graphs of \reffi{fi:aaff} contribute.
Let us first calculate the contribution $\M_{\mathrm{(a)}}$ of diagram (a) 
to the full helicity amplitude $\M$. We identify the photon
$\gamma(k,\lambda)$ in the subgraph shown in \reffi{fi:aff}a
with $\gamma(k_1,\lambda_1)$ and denote the contribution of this
subgraph by $T^{\gamma_1\Pf}_{\lambda_1}(\sigma)$. $\M_{\mathrm{(a)}}$ 
then follows upon multiplication of $T^{\gamma_1\Pf}_{\lambda_1}(\sigma)$, 
which is given in \refeq{eq:Taf}, with the second $\gamma\Pf\bar\Pf$ vertex 
and the wave functions for the second photon and the antifermion,
\beq
\ri\M_{\mathrm{(a)}}(\lambda_1,\lambda_2,\sigma,\sigma') = 
T^{\gamma_1\Pf}_{\lambda_1}(\sigma)(-\ri Q_\Pf e)
\pmatrix{ 0 & \delta^{\dot G}_{\dot F}\delta^H_E \cr
	  \epsilon^{\dot E\dot G}\epsilon^{FH} & 0 }
\pmatrix{ \vphantom{\phi_{\dot F}}\phi'_F \cr \psi^{\prime\dot F} }
\veps_{2,\dot GH}. 
\label{eq:aaffMa}
\eeq
Note that we follow the common practice to split off a
factor $+\ri$ from helicity amplitudes throughout. It is trivial to
carry out the spinor contractions in \refeq{eq:aaffMa} for the explicit 
insertions of $T^{\gamma_1\Pf}_{\lambda_1}(\sigma)$ and $\veps_{2,\dot GH}$.
The contribution $\M_{\mathrm{(b)}}$ of the diagram in \reffi{fi:aaff}b 
follows from $\M_{\mathrm{(a)}}$ by consistently interchanging the incoming 
photons. This leads us to the final result for the full helicity amplitudes
$\M=\M_{\mathrm{(a)}}+\M_{\mathrm{(b)}}$: 
\newcommand{\MPKi}{(p\cdot k_1)}
\newcommand{\MPKii}{(p\cdot k_2)}
\newcommand{\KiKii}   {\langle k_1 k_2 \rangle}
\newcommand{\CKiKii}  {\KiKii^*}
\newcommand{\phipsi}{\langle\phi\psi\rangle}
\newcommand{\PSIPHI}{\langle\psi'\phi'\rangle}
\newcommand{\phiPSI}{\langle\phi\psi'\rangle}
\newcommand{\PHIpsi}{\langle\phi'\psi\rangle}
\newcommand{\CphiPSI}{\phiPSI^*}
\newcommand{\CPHIpsi}{\PHIpsi^*}
\newcommand{\Kiiphi}  {\langle k_2 \phi \rangle}
\newcommand{\Kipsi}   {\langle k_1 \psi \rangle}
\newcommand{\KiPHI}   {\langle k_1 \phi' \rangle}
\newcommand{\KiiPSI}  {\langle k_2 \psi' \rangle}
\newcommand{\CKiiphi}  {\Kiiphi^*}
\newcommand{\CKipsi}   {\Kipsi^*}
\newcommand{\CKiPHI}   {\KiPHI^*}
\newcommand{\CKiiPSI}  {\KiiPSI^*}
\newcommand{\KiiPKi}  {\langle k_2 P k_1 \rangle}
\beqar
\M(+,+,\sigma,\sigma') &=&
\frac{Q_\Pf^2 e^2 \Mf}{2\MPKi\MPKii} \KiKii^2 \CphiPSI,
\nn\\
\M(-,-,\sigma,\sigma') &=&
\frac{Q_\Pf^2 e^2 \Mf}{2\MPKi\MPKii} (\CKiKii)^2 \PHIpsi,
\nn\\
\M(+,-,\sigma,\sigma') &=&
-\frac{Q_\Pf^2 e^2}{2\MPKi\MPKii} \KiiPKi
\Big( \KiPHI\CKiiphi+\Kipsi\CKiiPSI \Big),
\nn\\
\M(-,+,\sigma,\sigma') &=&
\M(+,-,\sigma,\sigma')\Big|_{k_1\leftrightarrow k_2}.
\label{eq:aaffres}
\eeqar
Here we have introduced the abbreviation 
\beq
\langle k_l P k_n \rangle = k_{l,\dot A} P^{\dot AB} k_{n,B} =
\sum_{i=1,2} \langle k_l \kappa_i \rangle^*\langle k_n \kappa_i \rangle,
\qquad l,n=1,2.
\label{eq:kPk}
\eeq
Moreover, we mention that Schouten's identity \refeq{eq:schouten} has
been used in order to get compact results for
$\lambda_1=\lambda_2=\pm$. 

As explained in \refse{se:discsym}, parity
relates amplitudes with opposite-helicity configurations 
through complex conjugation, modulo a sign factor. According to \refeq{eq:MP},
these relations explicitly read
\beq
\M(-\lambda_1,-\lambda_2,-\sigma,-\sigma') = 
-{\mathrm{sgn}}(\sigma\sigma')\M(\lambda_1,\lambda_2,\sigma,\sigma')^*,
\eeq
consistent with the above results. Alternatively, these relations
can be used to generate all helicity amplitudes from the generic results
for $\M(+,+,\sigma,\sigma')$ and $\M(+,-,\sigma,\sigma')$ given in
\refeq{eq:aaffres}. The process $\gamma\gamma\to f\bar f$ is also
CP-symmetric. Taking into account that the CP transformation interchanges
$f$ and $\bar f$, Eq.~\refeq{eq:MCP} implies
\beq
\M(-\lambda_1,-\lambda_2,-\sigma',-\sigma) = 
{\mathrm{sgn}}(\sigma\sigma')
\M(\lambda_1,\lambda_2,\sigma,\sigma')^* \Big|_{p\leftrightarrow p'},
\eeq
where the substitution $p\leftrightarrow p'$ also includes the
interchange of the respective spinors $\kappa_i$ and $\kappa'_i$.

\subsection{The process \boldmath{$f\bar f\to\gamma\gamma\gamma$}
and related reactions}

\begin{figure}
\setlength{\unitlength}{1pt}
\centerline{
\begin{picture}(300,80)(0,0)
\ArrowLine(10,70)(40,70)
\ArrowLine(40,70)(40,40)
\ArrowLine(40,40)(40,10)
\ArrowLine(40,10)(10,10)
\Photon   (40,10)(70,10){2}{4}
\Photon   (40,40)(70,40){2}{4}
\Photon   (40,70)(70,70){2}{4}
\Vertex   (40,70){2.0}
\Vertex   (40,40){2.0}
\Vertex   (40,10){2.0}
\Text     (100,40)[l]{+ \quad five permutations of the photons}
\end{picture} }
\caption{Born diagrams for $f\bar f\to\gamma\gamma\gamma$.}
\label{fi:ffaaa}
\end{figure}
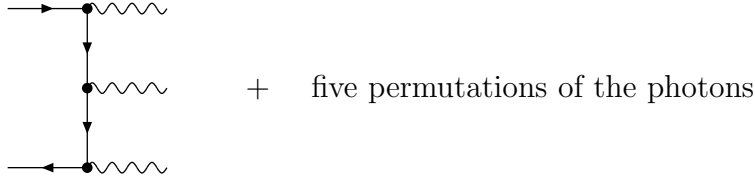

Next we consider fermion--antifermion annihilation into three photons:
\beq
f(p,\sigma) + \bar f(p',\sigma') \; \longrightarrow \;
\gamma(k_1,\lambda_1) + \gamma(k_2,\lambda_2) + \gamma(k_3,\lambda_3).
\eeq
The momentum matrices $P$ and $P'$ for the momenta $p$ and $p'$,
respectively, are again decomposed into the respective spinors 
$\kappa_{i,A}$ and $\kappa'_{i,A}$, as defined in \refeq{eq:pdecomp}
and \refeq{eq:ppdecomp}. The Dirac spinors are generically assigned by
\beq
\Psi^{(+)}_p = 
\pmatrix{ \vphantom{\phi_{\dot A}}\phi_A \cr \psi^{\dot A} }, 
\qquad
\overline\Psi^{(-)}_{p'} = (\psi^{\prime A},\phi'_{\dot A})
\label{eq:ffaaaPsi1}
\eeq
with the actual insertions
\beq
(\phi,\psi) = \Biggl\{
\barr{ll} (\kappa_1,-\kappa_2) & \quad \mbox{for} \; \sigma=+, \\
          (\kappa_2, \kappa_1) & \quad \mbox{for} \; \sigma=-, \earr
\qquad
(\phi',\psi') = \Biggl\{
\barr{ll} ( \kappa'_1,\kappa'_2) & \quad \mbox{for} \; \sigma'=-, \\
          (-\kappa'_2,\kappa'_1) & \quad \mbox{for} \; \sigma'=+. \earr
\label{eq:ffaaaPsi2}
\eeq
{}For all helicity configurations we have $\phipsi=\PSIPHI=\Mf$. The
polarization spinors $\veps^*_{\lambda_i,\dot AB}(k_i)$ for the outgoing
photons are defined as in \refeq{eq:m0polspinors}. Following the
strategy of \refse{se:gspinor} and ``crossing'' the results 
\refeq{eq:Taf} or \refeq{eq:Tafbar} for the 
subdiagrams of \reffi{fi:aff}, the actual calculation of the six diagrams for 
$f\bar f\to\gamma\gamma\gamma$ (see \reffi{fi:ffaaa}) is rather simple.
In order to minimize the number of generic amplitudes that have to be
calculated, we first give the relations that follow from discrete
symmetries. Parity, CP, and Bose symmetry imply
\beqar
\M(-\sigma,-\sigma',-\lambda_1,-\lambda_2,-\lambda_3) &=&
-\mathrm{sgn}(\sigma\sigma')
\M(\sigma,\sigma',\lambda_1,\lambda_2,\lambda_3)^*,
\nn\\[.5em]
\M(-\sigma',-\sigma,-\lambda_1,-\lambda_2,-\lambda_3) &=&
-{\mathrm{sgn}}(\sigma\sigma')
\M(\sigma,\sigma',\lambda_1,\lambda_2,\lambda_3)^* 
\Big|_{p\leftrightarrow p'},
\nn\\[.5em]
\M(\sigma,\sigma',\lambda_{i_1},\lambda_{i_2},\lambda_{i_3})
&=&
\M(\sigma,\sigma',\lambda_1,\lambda_2,\lambda_3) 
\Big|_{k_1\to k_{i_1},k_2\to k_{i_2},k_3\to k_{i_3}}.
\label{eq:ffaaaPB}
\eeqar
Therefore, for any given $(\sigma,\sigma')$ combination it suffices to 
calculate the helicity amplitudes with 
$(\lambda_1,\lambda_2,\lambda_3)=({+}{+}{+}),({+}{+}{-})$, from which all 
other amplitudes follow by \refeq{eq:ffaaaPB}. These generic results can be
expressed in a very compact form:

\newcommand{\MPKiii}{(p\cdot k_3)}
\newcommand{\MPpKiii}{(p'\cdot k_3)}
\newcommand{\MPpKi}{(p'\cdot k_1)}
\newcommand{\MPpKii}{(p'\cdot k_2)}
\newcommand{\KiKiii}  {\langle k_1 k_3 \rangle}
\newcommand{\KiiKiii} {\langle k_2 k_3 \rangle}
\newcommand{\CKiKiii} {\KiKiii^*}
\newcommand{\CKiiKiii}{\KiiKiii^*}
\newcommand{\KiPKii}  {\langle k_1 P k_2 \rangle}
\newcommand{\KiPKiii} {\langle k_1 P k_3 \rangle}
\newcommand{\KiiPKiii}{\langle k_2 P k_3 \rangle}
\newcommand{\KiiiPKi} {\langle k_3 P k_1 \rangle}
\newcommand{\KiiiPKii}{\langle k_3 P k_2 \rangle}
\newcommand{\KiPpPKi}     {\langle k_1 P' P k_1 \rangle}
\newcommand{\KiiPpPKii}   {\langle k_2 P' P k_2 \rangle}
\newcommand{\KiiiPpPKiii} {\langle k_3 P' P k_3 \rangle}
\newcommand{\CKiPpPKi}    {\KiPpPKi^*}
\newcommand{\CKiiPpPKii}  {\KiiPpPKii^*}
\newcommand{\CKiiiPpPKiii}{\KiiiPpPKiii^*}
\newcommand{\Kiphi}   {\langle k_1 \phi \rangle}
\newcommand{\Kiiiphi} {\langle k_3 \phi \rangle}
\newcommand{\Kiipsi}  {\langle k_2 \psi \rangle}
\newcommand{\Kiiipsi} {\langle k_3 \psi \rangle}
\newcommand{\KiiPHI}  {\langle k_2 \phi' \rangle}
\newcommand{\KiiiPHI} {\langle k_3 \phi' \rangle}
\newcommand{\KiPSI}   {\langle k_1 \psi' \rangle}
\newcommand{\KiiiPSI} {\langle k_3 \psi' \rangle}
\newcommand{\CKiphi}   {\Kiphi^*}
\newcommand{\CKiiiphi} {\Kiiiphi^*}
\newcommand{\CKiipsi}  {\Kiipsi^*}
\newcommand{\CKiiipsi} {\Kiiipsi^*}
\newcommand{\CKiiPHI}  {\KiiPHI^*} 
\newcommand{\CKiiiPHI} {\KiiiPHI^*}
\newcommand{\CKiPSI}   {\KiPSI^*}
\newcommand{\CKiiiPSI} {\KiiiPSI^*}
\beq
\M(\sigma,\sigma',\lambda_1,\lambda_2,\lambda_3) =
\frac{Q_f^3e^3A_{\lambda_1\lambda_2\lambda_3}(\sigma,\sigma')}
{4\sqrt{2}\MPKi\MPKii\MPKiii}, 
\eeq
where
\beqar
A_{{+}{+}{+}}(\sigma,\sigma') &=& \Mf \frac{(\CKiiKiii)^2}{\MPpKi}
\KiPpPKi \phiPSI \;+\; \mbox{cyclic permutations in } k_1,k_2,k_3,
\nn\\
A_{{+}{+}{-}}(\sigma,\sigma') &=& \Mf\frac{(\CKiKii)^2}{2\MPpKiii}
\Big( \CKiiiPpPKiii\phiPSI+2\MPKiii\Kiiiphi\KiiiPSI \Big)
\nn\\
&& {} -\frac{\KiiPKiii}{\MPpKi} \Big[ 
\KiPpPKi \Big( \CKiipsi\KiiiPSI+\CKiiPHI\Kiiiphi \Big)
\nn\\
&& {} \phantom{ +\frac{\KiiPKiii}{\MPpKi} \Big[ }
+2\MPKi\Kiiiphi\CKiPHI\CKiKii \Big]
\;+\; (k_1\leftrightarrow k_2).
\eeqar
In addition to the shorthand $\langle k_l P k_n \rangle$ of \refeq{eq:kPk}
we have introduced the useful abbreviations
\beq
\langle k_l P' P k_l \rangle = 
k_{l,\dot A} P^{\prime\dot AB} P_{\dot CB} k^{\dot C}_l = \sum_{i,j=1,2} 
\langle k_l \kappa'_i \rangle^* \langle \kappa_j \kappa'_i \rangle
\langle \kappa_j k_l \rangle^*,
\qquad l=1,2,3.
\label{eq:kPpPk}
\eeq

The above results for $f\bar f\to\gamma\gamma\gamma$ can also be used
to generate the helicity amplitudes for the bremsstrahlung processes
$\gamma\gamma\to\Pf\bar\Pf\gamma$ and $\Pem\gamma\to\Pem\gamma\gamma$,
if the crossing relations of \refse{se:discsym} are applied.
In particular, we get a set of helicity amplitudes for the so-called double 
Compton process
\beq
\Pem(p,\sigma) + \gamma(k,\lambda) \; \longrightarrow \;
\Pem(p',\sigma') + \gamma(k'_1,\lambda'_1) + \gamma(k'_2,\lambda'_2)
\label{eq:eaeaa}
\eeq
by identifying $f=\Pem$, consistently replacing
\beq
\begin{array}[b]{rlrlrlrl}
k_3^\mu &\to -k^\mu, \qquad & k_{3,A} & \to k_{3,A}, \qquad &
k_{3,\dot A} & \to -k_{3,\dot A}, \qquad & \lambda_3 & \to -\lambda,
\\
p^{\prime\mu} &\to -p^{\prime\mu}, \qquad & \kappa'_{i,A} & \to\kappa'_{i,A}, 
\qquad &
\kappa'_{i,\dot A} & \to -\kappa'_{i,\dot A}, \qquad & \sigma' & \to -\sigma'
\end{array}
\eeq
in all expressions, and adding the factor ${\mathrm{sgn}}(\sigma')$,
where $\sigma'$ is the helicity of the outgoing $\Pem$ in \refeq{eq:eaeaa}. 
{}Finally, we mention that
our helicity amplitudes have been numerically reproduced (up to phase
factors) in the framework of the calculation presented in \citere{de98}
by employing the spinor method of \citere{sp2} and by a third, completely
different method.

\subsection{The process \boldmath{$\mu^-\mu^+\to f\bar f\gamma$} 
and related reactions}

\begin{figure}
\setlength{\unitlength}{1pt}
\centerline{
\begin{picture}(120,80)(0,0)
\ArrowLine(10,70)(25,55)
\ArrowLine(25,55)(40,40)
\ArrowLine(40,40)(10,10)
\Photon   (40,40)(80,40){2}{4}
\ArrowLine(80,40)(110,70)
\ArrowLine(110,10)(80,40)
\Photon   (25,55)(60,70){2}{4}
\Vertex   (40,40){2.0}
\Vertex   (80,40){2.0}
\Vertex   (25,55){2.0}
\Text     ( -8,70)[l]{$\mu^-$}
\Text     ( -8,10)[l]{$\mu^+$}
\Text     (115,10)[l]{$\bar f$}
\Text     (115,70)[l]{$f$}
\Text     ( 68,70)[l]{$\gamma$}
\Text     ( 50,28)[l]{$\gamma,Z$}
\end{picture} 
\hspace*{2em}
\begin{picture}(150,80)(0,0)
\ArrowLine(10,70)(25,55)
\ArrowLine(25,55)(40,40)
\ArrowLine(40,40)(10,10)
\DashLine(40,40)(80,40){5}
\ArrowLine(80,40)(110,70)
\ArrowLine(110,10)(80,40)
\Photon   (25,55)(60,70){2}{4}
\Vertex   (40,40){2.0}
\Vertex   (80,40){2.0}
\Vertex   (25,55){2.0}
\Text     ( -8,70)[l]{$\mu^-$}
\Text     ( -8,10)[l]{$\mu^+$}
\Text     (115,10)[l]{$\bar f$}
\Text     (115,70)[l]{$f$}
\Text     ( 68,70)[l]{$\gamma$}
\Text     ( 50,28)[l]{$\chi,H$}
\Text     (135,40)[l]{etc.}
\end{picture} } 
\caption{Born diagrams for $\mu^-\mu^+\to f\bar f\gamma$, where graphs with
the outgoing photon attached to the other charged fields are suppressed.}
\label{fi:mumuffa}
\end{figure}
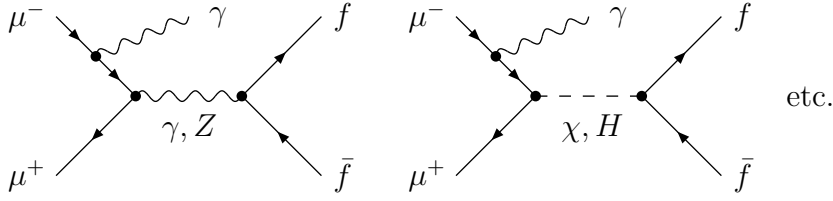

As a final example, we consider the process
\beq
\mu^-(p,\sigma) + \mu^+(p',\sigma') \;\longrightarrow\;
f(q,\tau) + \bar f(q',\tau') + \gamma(k,\lambda),
\eeq
which will be of relevance at future muon colliders.
The momenta $p$, $p'$ and the
corresponding Dirac spinors $\Psi^{(+)}_p$, 
$\overline\Psi^{(-)}_{p'}$ for the incoming muons are defined in the
same way as in the last section [see \refeq{eq:ffaaaPsi1} and
\refeq{eq:ffaaaPsi2}]. The momentum matrices $Q$ and $Q'$ for the
outgoing momenta $q$ and $q'$ are decomposed into the spinors
$\rho_{i,A}$ and $\rho'_{i,A}$, respectively,
\beq
Q_{\dot AB} = \sum_{i=1,2} \rho_{i,\dot A} \rho_{i,B}, \qquad
Q'_{\dot AB} = \sum_{i=1,2} \rho'_{i,\dot A} \rho'_{i,B}.
\eeq
The respective Dirac spinors are generically given by
\beq
\overline\Psi^{(+)}_q = (\eta^A,\xi_{\dot A}),
\qquad
\Psi^{(-)}_{q'} = 
\pmatrix{ \vphantom{\xi'_{\dot A}}\xi'_A \cr \eta^{\prime\dot A} }
\eeq
with the actual insertions
\newcommand{\xieta}{\langle\xi  \eta\rangle}
\newcommand{\ETAXI}{\langle\eta'\xi'\rangle}
\beq
(\xi,\eta) = \Biggl\{
\barr{ll} (\rho_1,-\rho_2) & \quad \mbox{for} \; \tau=+, \\
          (\rho_2, \rho_1) & \quad \mbox{for} \; \tau=-, \earr
\qquad
(\xi',\eta') = \Biggl\{
\barr{ll} ( \rho'_1,\rho'_2) & \quad \mbox{for} \; \tau'=-, \\
          (-\rho'_2,\rho'_1) & \quad \mbox{for} \; \tau'=+, \earr
\eeq
i.e.\ we have $\xieta=\ETAXI=\Mf$. The
polarization spinors $\veps^*_{\lambda,\dot AB}(k)$ for the outgoing
photon are defined as in \refeq{eq:m0polspinors}. For $f\ne\mu^-$ 16
diagrams contribute to the process in the Standard Model at tree level;
they are schematically indicated in \reffi{fi:mumuffa}.
The case $f=\mu^-$ will not be considered here explicitly; the results
for this particular channel follow from the ones for $f\ne\mu^-$ by
adding the amplitudes for the crossed process $f\mu^+\to f\mu^+\gamma$
with negative sign. Since the interaction is mediated by the exchange of
neutral particles only, the electromagnetic currents of the muon and 
the fermion $f$ are conserved separately. This means that the gauge
spinor for the outgoing photon can be chosen differently for initial and
final-state radiation, which simplifies the calculation. In the
following we generically denote the couplings of the neutral bosons to 
the fermions by
\beqar
g_{\gamma f}^\pm &=& -Q_f, \qquad
g_{\PZ f}^+ = -\frac{\sw}{\cw}Q_f, \qquad
g_{\PZ f}^- = -\frac{\sw}{\cw}Q_f + \frac{I_{W,f}}{\cw\sw},
\nn\\
g_{\PH f}^\pm &=& -\frac{1}{2\sw}\frac{\Mf}{\MW}, \qquad
g_{\chi f}^\pm = \pm\frac{\ri I_{W,f}}{\sw}\frac{\Mf}{\MW},
\label{eq:couplings}
\eeqar
where $I_{W,f}=\pm\frac{1}{2}$ is the weak isospin of the left-handed
part of the fermion field $f$. In \refeq{eq:couplings} we follow the
conventions of \citere{sm} for the Standard Model parameters and fields;
in particular, $\chi$ denotes the would-be Goldstone partner to the
Z~boson, and H is the physical Higgs boson. 
Since Standard Electroweak Theory does
not conserve parity, the parity-induced relations between helicity
amplitudes with opposite helicity configurations involve also an
interchange of the chiral couplings:
\beq
\M(-\sigma,-\sigma',-\tau,-\tau',-\lambda) =
{\mathrm{sgn}}(\sigma\sigma'\tau\tau') \M(\sigma,\sigma',\tau,\tau',\lambda)^*
\Big|_{ (g_{\dots}^\pm)^* \leftrightarrow g_{\dots}^\mp }.
\label{eq:mumuffaP}
\eeq
On the other hand, the process is CP-symmetric at the considered
perturbative level, and the relations induced by CP symmetry read
\beq
\M(-\sigma',-\sigma,-\tau',-\tau,-\lambda) =
-{\mathrm{sgn}}(\sigma\sigma'\tau\tau') \M(\sigma,\sigma',\tau,\tau',\lambda)^*
\Big|_{p\leftrightarrow p', q\leftrightarrow q'}.
\label{eq:mumuffaCP}
\eeq
The matrix elements $\M$ are decomposed according to the boson in the 
$s$-channel, 
\beq
\M(\sigma,\sigma',\tau,\tau',\lambda) = 
\sqrt{2}e^3 \Biggl[ 
\sum_{\PV=\gamma,\PZ} A^{(\PV)}_\lambda(\sigma,\sigma',\tau,\tau') 
+ \sum_{\PS=\chi,\PH} A^{(\PS)}_\lambda(\sigma,\sigma',\tau,\tau') \Biggr],
\label{eq:mumuffaamp}
\eeq
leading to two generic functions $A^{(\PV)}_\lambda$ and
$A^{(\PS)}_\lambda$ for the exchange of a vector field and a scalar
field, respectively. Owing to \refeq{eq:mumuffaP} or
\refeq{eq:mumuffaCP}, it is sufficient to
give the results for \mbox{$\lambda=+$},

\newcommand{\gVfp}{g_{\PV f}^+}
\newcommand{\gVfm}{g_{\PV f}^-}
\newcommand{\gVmup}{g_{\PV\mu}^+}
\newcommand{\gVmum}{g_{\PV\mu}^-}
\newcommand{\gSfp}{g_{\PS f}^+}
\newcommand{\gSfm}{g_{\PS f}^-}
\newcommand{\gSmup}{g_{\PS\mu}^+}
\newcommand{\gSmum}{g_{\PS\mu}^-}
\newcommand{\KPpPK}     {\langle k P' P k \rangle}
\newcommand{\KQpQK}     {\langle k Q' Q k \rangle}
\newcommand{\MPpK}{(p'\cdot k)}
\newcommand{\MQK}{(q\cdot k)}
\newcommand{\MQpK}{(q'\cdot k)}
\newcommand{\phiXI} {\langle \phi  \xi'  \rangle}
\newcommand{\phieta}{\langle \phi  \eta  \rangle}
\newcommand{\psixi} {\langle \psi  \xi   \rangle}
\newcommand{\psiETA}{\langle \psi  \eta' \rangle}
\newcommand{\PHIxi} {\langle \phi' \xi   \rangle}
\newcommand{\PHIETA}{\langle \phi' \eta' \rangle}
\newcommand{\PSIXI} {\langle \psi' \xi'  \rangle}
\newcommand{\PSIeta}{\langle \psi' \eta  \rangle}
\newcommand{\XIeta} {\langle \xi'  \eta  \rangle}
\newcommand{\xiETA} {\langle \xi   \eta' \rangle}
\newcommand{\CphiXI} {\phiXI^*}
\newcommand{\Cphieta}{\phieta^*}
\newcommand{\Cpsixi} {\psixi^*}
\newcommand{\CpsiETA}{\psiETA^*}
\newcommand{\CXIeta} {\XIeta^*}
\newcommand{\CxiETA} {\xiETA^*}
\newcommand{\CPHIxi} {\PHIxi^*}
\newcommand{\CPHIETA}{\PHIETA^*}
\newcommand{\CPSIXI} {\PSIXI^*}
\newcommand{\CPSIeta}{\PSIeta^*}
\newcommand{\Kpsi}{\langle k \psi \rangle}
\newcommand{\Kxi} {\langle k \xi \rangle}
\newcommand{\KPHI}{\langle k \phi' \rangle}
\newcommand{\KXI} {\langle k \xi' \rangle}
\newcommand{\KETA}{\langle k \eta' \rangle}
\newcommand{\CKpsi}{\Kpsi^*}
\newcommand{\CKxi} {\Kxi^*}
\newcommand{\CKPHI}{\KPHI^*}
\newcommand{\CKXI} {\KXI^*}
\newcommand{\CKETA}{\KETA^*}
\beqar
A^{(\PV)}_+(\sigma,\sigma',\tau,\tau') &=&
\biggl\{ \frac{Q_\mu\KPpPK}{2\MPK\MPpK[(q+q')^2-\MV^2]}
- \frac{Q_f\KQpQK}{2\MQK\MQpK[(p+p')^2-\MV^2]} \biggr\}
\nn\\ 
&& {} \quad \times \Bigl(
\gVmup\gVfp\CPHIxi\phiXI + \gVmup\gVfm\CPHIETA\phieta 
\nn\\ 
&& \phantom{{} \quad \times \Bigl(}
+ \gVmum\gVfp\Cpsixi\PSIXI + \gVmum\gVfm\CpsiETA\PSIeta \Big)
\nn\\ 
&& {} - \frac{Q_\mu}{(q+q')^2-\MV^2} \biggl[ 
\frac{\gVmup\CKPHI}{\MPpK} \Big( \gVfp\CKxi\phiXI+\gVfm\CKETA\phieta \Big)
\nn\\ 
&& \phantom{{} - \frac{Q_\mu}{(q+q')^2-\MV^2} \biggl[ }
- \frac{\gVmum\CKpsi}{\MPK} \Big( \gVfp\CKxi\PSIXI+\gVfm\CKETA\PSIeta \Big) 
\biggr]
\nn\\ 
&& {} - \frac{Q_f}{(p+p')^2-\MV^2} \biggl[ 
\frac{\gVfp\CKxi}{\MQK} \Big( \gVmup\CKPHI\phiXI+\gVmum\CKpsi\PSIXI \Big)
\nn\\ 
&& \phantom{{} - \frac{Q_f}{(p+p')^2-\MV^2} \biggl[ }
- \frac{\gVfm\CKETA}{\MQpK} \Big( \gVmup\CKPHI\phieta+\gVmum\CKpsi\PSIeta \Big)
\biggr],
\nn\\[.5em]
A^{(\PS)}_+(\sigma,\sigma',\tau,\tau') &=&
\biggl\{ \frac{Q_f\KQpQK}{4\MQK\MQpK[(p+p')^2-\MS^2]} 
- \frac{Q_\mu\KPpPK}{4\MPK\MPpK[(q+q')^2-\MS^2]} \biggr\}
\nn\\ 
&& {} \quad \times \Bigl( \gSmup\phiPSI+\gSmum\CPHIpsi \Big)
\Bigl( \gSfp\XIeta+\gSfm\CxiETA \Big)
\nn\\ 
&& {} + \frac{Q_\mu\gSmum\CKPHI\CKpsi}{(q+q')^2-\MS^2} 
\Bigl( \gSfp\XIeta+\gSfm\CxiETA \Big)
\biggl[ \frac{1}{2\MPK}+\frac{1}{2\MPpK} \biggr]
\nn\\ 
&& {} - \frac{Q_f\gSfm\CKxi\CKETA}{(p+p')^2-\MS^2} 
\Bigl( \gSmup\phiPSI+\gSmum\CPHIpsi \Big)
\biggl[ \frac{1}{2\MQK}+\frac{1}{2\MQpK} \biggr].
\nn\\
\eeqar
The quantities $\KPpPK$ and $\KQpQK$ are defined in complete analogy to
$\langle k_l P' P k_l \rangle$ in \refeq{eq:kPpPk}.
The results for $A^{(\PV)}_\lambda$ and $A^{(\PS)}_\lambda$ 
are written down in the 't~Hooft--Feynman gauge, i.e.\ we have
$M_\chi=\MZ$. We mention, however, that we have reproduced the result
for the complete matrix element $\M$
also in an arbitrary $R_\xi$-gauge for the photon and Z-boson fields, in
which the individual contributions $A^{(\PV)}_\lambda$ and
$A^{(\PS)}_\lambda$ look different.

The above results have also been used to derive the helicity amplitudes
for the process
\beq
\Pem(p,\sigma) + \gamma(k,\lambda) \; \longrightarrow \;
\Pem(p',\sigma') + \Pem(q,\tau) + \Pep(q',\tau')
\label{eq:eaeee}
\eeq
in QED. The QED part of the amplitudes \refeq{eq:mumuffaamp} is obtained
by taking the contribution of $\PV=\gamma$ in \refeq{eq:mumuffaamp} only.
Moreover, we identify $Q_\mu=Q_f=Q_\Pe=-1$.
The crossing proceeds analogously to the case described at the end of
the previous section, i.e.\ one has to perform the replacements
\beq
\begin{array}[b]{rlrlrlrl}
k^\mu &\to -k^\mu, \qquad & k_A & \to k_A, \qquad &  
k_{\dot A} & \to -k_{\dot A}, \qquad & \lambda & \to -\lambda,
\\
p^{\prime\mu} &\to -p^{\prime\mu}, \qquad & \kappa'_{i,A} & \to\kappa'_{i,A}, 
\qquad &
\kappa'_{i,\dot A} & \to -\kappa'_{i,\dot A}, \qquad & \sigma' & \to -\sigma'
\end{array}
\eeq
and to apply the sign factor
${\mathrm{sgn}}(\sigma')$, where $\sigma'$ is the helicity of the
outgoing $\Pem$ in \refeq{eq:eaeee}. Finally,
we have to antisymmetrize all amplitudes with respect to the
interchange of the two outgoing electrons. 
Moreover, the amplitudes have been numerically
reproduced by an independent helicity method in the framework of the 
calculation discussed in \citere{de98}.
The polarized cross sections of 
$\Pem\gamma\to\Pem\Pem\Pep$ contribute, in particular, to the left--right 
asymmetry $A_{\mathrm{LR}}$ of polarized Compton scattering as background. 
The numerical agreement of the results presented in \citere{de98} for this 
contribution to $A_{\mathrm{LR}}$ with the completely independent ones
of \citere{sw97} represents an additional check of the calculation. 

\section{Summary}
\label{se:sum}

The Weyl--van-der-Waerden (WvdW) spinor technique for the calculation of
helicity amplitudes of massive and massless particles
has been described in detail,
providing all necessary ingredients for an implementation in computer
algebra. This formalism leads to rather compact results for amplitudes, 
which can be directly used for numerical evaluations, immediately after 
all spinor indices have been contracted to so-called spinor products. 
These contractions are technically similar to the usual ones for Lorentz 
indices. Since all mathematical objects, such as momenta, polarization 
vectors, and Dirac
spinors, are expressed in terms of WvdW spinors, the spinor calculus
often allows for further simplifications at the analytical level.
Moreover, we have formulated how to exploit discrete symmetries for a
reduction of the algebraic work or for providing additional checks, and
how to avoid problems that are due to the appearance of unphysical poles 
in amplitudes.

The use and the power of the described spinor technique have been 
demonstrated by the explicit calculation of the helicity amplitudes for
the processes $\gamma\gamma\to\Pf\bar\Pf$, 
$\Pf\bar\Pf\to\gamma\gamma\gamma$, $\mu^-\mu^+\to\Pf\bar\Pf\gamma$ with
massive fermions, and of
reactions that can be obtained from those by crossing symmetry. The
results, which have been analytically simplified as far as possible, are 
very compact and well suited to numerical evaluations.

\appendix
\def\thesection{A}

\section*{Appendix}

\section*{More details about discrete symmetries}

In this appendix we supplement the discussion of parity and CP symmetry
in \refse{se:discsym} by further details.

\begin{enumerate}
\item Parity: \\
In order to explicitly derive the relations between helicity amplitudes
that are connected by parity, we inspect the individual
terms in an amplitude $\M$ after complex conjugation. Contractions between
momentum matrices, which are just Minkowski inner products according to
\refeq{eq:kp}, are not changed at all, since the result is real.
Contractions between a momentum $k$ and a polarization vector $\veps$, 
or between two polarization vectors, turn into the products for the 
corresponding polarization vectors with opposite helicity, since
$(K_{\dot AB})^*=K_{A\dot B}=K_{\dot BA}$ and
$(\veps_{\lambda,{\dot A}B})^*=\veps_{\lambda,A{\dot B}}=
\veps_{-\lambda,{\dot B}A}\,$ ($\lambda=0,\pm 1$).
Coupling factors $\ri C$ for interactions between bosons simply turn into
$-\ri C^*$, and boson propagators receive a factor $-1$ owing to the
explicit factor $\ri$ in the numerator.
The only non-trivial terms are those originating from Dirac matrices and
fermionic wave functions. The complex conjugate amplitude remains unchanged 
if we multiply each fermionic vertex by a factor 
$\Gamma_{\mathrm{P}}^2=-{\bf 1}$ from the left and from the right, 
where the matrix $\Gamma_{\mathrm{P}}$ is defined as
\beq
\Gamma_{\mathrm{P}} = \ri\gamma^2\gamma^5 = 
\pmatrix{0 & \epsilon \cr \epsilon & 0}.
\eeq
Shifting then a factor $\Gamma_{\mathrm{P}}$ to the fermion propagators
or fermionic wave functions that are attached on the right side of this vertex,
and shifting a factor $-\Gamma_{\mathrm{P}}$ to the left side, we get the
following replacements when going over from $\M$ to $\M^*$:
\beqar
\ri e\ks C^\tau_{V\bar f_1f_2}\omega_\tau & \to &
-\Gamma_{\mathrm{P}} \left[\ri e\ks C^\tau_{V\bar f_1f_2}\omega_\tau \right]^*
\Gamma_{\mathrm{P}} = 
-\ri e\ks \left(C^{-\tau}_{V\bar f_1f_2}\right)^*\omega_\tau, 
\nn\\ {}
\ri e\es_\lambda C^\tau_{V\bar f_1f_2}\omega_\tau & \to &
-\Gamma_{\mathrm{P}} 
\left[\ri e\es_\lambda C^\tau_{V\bar f_1f_2}\omega_\tau \right]^*
\Gamma_{\mathrm{P}} = 
-\ri e\es_{-\lambda} \left(C^{-\tau}_{V\bar f_1f_2}\right)^*\omega_\tau, 
\nn\\ {}
\ri e C^\tau_{S\bar f_1f_2}\omega_\tau & \to &
-\Gamma_{\mathrm{P}} \left[\ri e C^\tau_{S\bar f_1f_2}\omega_\tau \right]^*
\Gamma_{\mathrm{P}} = 
-\ri e \left(C^{-\tau}_{S\bar f_1f_2}\right)^*\omega_\tau, 
\nn\\ {}
\ri(\slash k+m_f) & \to &
\Gamma_{\mathrm{P}} \left[\ri(\slash k+m_f)\right]^* (-\Gamma_{\mathrm{P}}) = 
-\ri(\slash k+m_f),
\nn\\ {}
\Psi^{(\pm)}_{k,1} & \to &
\Gamma_{\mathrm{P}} \left[\Psi^{(\pm)}_{k,1}\right]^* = +\Psi^{(\pm)}_{k,2},
\qquad
\overline\Psi^{(\pm)}_{k,1} \to 
\left[\overline\Psi^{(\pm)}_{k,1}\right]^* (-\Gamma_{\mathrm{P}}) =
+\overline\Psi^{(\pm)}_{k,2},
\nn\\ {}
\Psi^{(\pm)}_{k,2} & \to &
\Gamma_{\mathrm{P}} \left[\Psi^{(\pm)}_{k,2}\right]^* = -\Psi^{(\pm)}_{k,1},
\qquad
\overline\Psi^{(\pm)}_{k,2} \to 
\left[\overline\Psi^{(\pm)}_{k,2}\right]^* (-\Gamma_{\mathrm{P}}) =
-\overline\Psi^{(\pm)}_{k,1},
\hspace*{3em}
\label{eq:parity}
\eeqar
\begin{sloppypar}
where the notation for the generic Feynman rules of \reftas{ta:exproprul} and
\ref{ta:vertrul} is used. 
The relations between $\M(\sigma_i;\lambda_j)^*$ and
$\M(-\sigma_i;-\lambda_j)$ can be read off from \refeq{eq:parity} and 
the above considerations. The substitutions for the
chiral couplings is explicitly given; in particular, there is no
substitution in parity-conserving theories, in which
$C^\tau_{V\bar f_1f_2} = \left(C^{-\tau}_{V\bar f_1f_2}\right)^*$ etc.
There is a global change of sign due to vertices and propagators,
since each of them introduces a factor $-1$, and since their total
number is always odd in tree-level amplitudes. Another factor $-1$
comes from our convention of extracting the factor $\ri$ from the
amplitude $\ri\M$.
{}Finally, we encounter a factor $\sgn(\sigma_i)$ for each external
fermion with helicity $\sigma_i$ and a factor $-\sgn(\sigma_i)$ for 
each external antifermion, leading us directly to relation
\refeq{eq:MP}.
\end{sloppypar}
\item CP symmetry: \\
As in the case of parity, we derive the CP-induced relations between 
helicity amplitudes by considering the complex conjugate of a given 
amplitude $\M$. We assume an underlying model in which all couplings
between bosons are CP-conserving, as it is the case in the Electroweak
Standard Model. Then, 
for bosonic fields the situation is simple, because the wave functions 
of particles and antiparticles are formally identical, and the bosonic 
couplings that are related by charge conjugation differ by at most the 
sign factors (see e.g.\ the Feynman rules in \citere{sm}).
Specifically, there is a factor $(-1)^v$ between each 
vertex factor $\ri C$ and the expression $\ri\bar C$ for the 
vertex involving the corresponding charge-conjugated fields, where
$v$ is the number of vector bosons attached to the vertex. 
In order to get rid of the factor $(-1)^v$, we
shift a factor $-1$ for each vector-boson leg of a vertex to the
attached vector-boson propagators or external wave functions, resulting
in an overall factor $(-1)^{n_V}$ for each diagram, where $n_V$ is the
number of external vector bosons. After this procedure the bosonic 
propagators and vertices for the charge-conjugated fields consistently 
differ by a factor $-1$ from the complex-conjugated counterparts
involving the original fields.

{}For fermionic fields we proceed in a way 
similar to the treatment of parity described
above. In addition to taking the complex conjugate of a Dirac chain, we
now also transpose each term and invert the order in the chain, 
since charge conjugation reverses the directions of fermionic lines.
Then we multiply fermionic vertices, propagators, and wave functions 
by the matrix
\beq
\Gamma_{\mathrm{CP}} = -\gamma^0\gamma^5 = 
\pmatrix{0 & {\bf 1} \cr -{\bf 1} & 0}
\eeq
with $\Gamma_{\mathrm{CP}}^2=-{\bf 1}$ in the same way as done with
$\Gamma_{\mathrm{P}}$ above. This leads to the replacements
\beqar
\ri e\ks C^\tau_{V\bar f_1f_2}\omega_\tau & \to &
-\Gamma_{\mathrm{CP}}
\left[-\ri e\ks C^\tau_{V\bar f_1f_2}\omega_\tau \right]^\dagger 
\Gamma_{\mathrm{CP}} =
-\ri e\ks \left(C^\tau_{V\bar f_1f_2}\right)^*\omega_\tau, 
\nn\\ {}
\ri e\es_\lambda C^\tau_{V\bar f_1f_2}\omega_\tau & \to &
-\Gamma_{\mathrm{CP}}
\left[-\ri e\es_\lambda C^\tau_{V\bar f_1f_2}\omega_\tau \right]^\dagger
\Gamma_{\mathrm{CP}} = 
-\ri e\es_{-\lambda} \left(C^\tau_{V\bar f_1f_2}\right)^*\omega_\tau, 
\nn\\ {}
\ri e C^\tau_{S\bar f_1f_2}\omega_\tau & \to &
-\Gamma_{\mathrm{CP}}
\left[\ri e C^\tau_{S\bar f_1f_2}\omega_\tau \right]^\dagger 
\Gamma_{\mathrm{CP}} = 
-\ri e \left(C^{-\tau}_{S\bar f_1f_2}\right)^*\omega_\tau, 
\nn\\ {}
\ri(\slash k+m_f) & \to &
\Gamma_{\mathrm{CP}} 
\left[\ri(\slash k+m_f)\right]^\dagger (-\Gamma_{\mathrm{CP}}) = 
-\ri(-\slash k+m_f),
\nn\\ {}
\Psi^{(\pm)}_{k,1} & \to &
\left[\Psi^{(\pm)}_{k,1}\right]^\dagger 
(-\Gamma_{\mathrm{CP}}) =
-\overline\Psi^{(\mp)}_{k,1},
\qquad
\overline\Psi^{(\pm)}_{k,1} \to 
\Gamma_{\mathrm{CP}}
\left[\overline\Psi^{(\pm)}_{k,1}\right]^\dagger =
+\Psi^{(\mp)}_{k,1},
\nn\\
\Psi^{(\pm)}_{k,2} & \to &
\left[\Psi^{(\pm)}_{k,2}\right]^\dagger 
(-\Gamma_{\mathrm{CP}}) =
+\overline\Psi^{(\mp)}_{k,2},
\qquad
\overline\Psi^{(\pm)}_{k,2} \to 
\Gamma_{\mathrm{CP}}
\left[\overline\Psi^{(\pm)}_{k,2}\right]^\dagger =
-\Psi^{(\mp)}_{k,2}.
\hspace*{3em}
\label{eq:cp}
\eeqar
Note that we had to include a factor $-1$ for the
vector-boson--fermion coupling in the first two lines, so as to be
consistent with the above treatment of bosonic couplings. Moreover, one
should realize the correct change of sign in the momentum of the
fermion propagator [see fourth line in \refeq{eq:cp}], 
corresponding to the inversion of the fermion line. In the case of
CP-violation the necessary substitutions for the couplings can be read
off from \refeq{eq:cp}; if CP is conserved no substitution is 
necessary, because then 
$C^\tau_{V\bar f_1f_2}=\left(C^\tau_{V\bar f_1f_2}\right)^*$ and
$C^\tau_{S\bar f_1f_2}=\left(C^{-\tau}_{S\bar f_1f_2}\right)^*$.
The overall sign between the amplitude
$\M(\sigma_i;\lambda_j)^*$ and the one for the CP-related process
$\overline{\M}(-\sigma_i;-\lambda_j)$ is deduced as follows. At tree
level, there is an overall factor $-1$ from the vertices and propagators
and another $-1$ from the convention for $\M$, as in the case of parity.
{}From \refeq{eq:cp} one
can see that incoming fermions and antifermions with helicity
$\sigma_i$ introduce each a factor $-\sgn(\sigma_i)$, while outgoing
fermionic lines yield a factor $\sgn(\sigma_i)$. Taking into account the
factor $(-1)^{n_V}$ derived above, we obtain relation \refeq{eq:MCP}.
\end{enumerate}

\section*{Acknowledgements}

The author thanks D.~de~Florian, A.~Denner and W.~Vogelsang for a
critical reading of the manuscript and for their
aid in making the presentation as transparent as possible.

\def\vol#1{{\bf #1}}
\def\mag#1{{\sl #1}}

\end{document}